\documentclass{aa}  

\usepackage{graphicx}
\usepackage{txfonts}
\usepackage{lipsum}
\usepackage{subcaption}         
\usepackage{lscape}             
\usepackage{placeins}           
\usepackage{natbib}

\bibpunct{(}{)}{;}{a}{}{,}
\usepackage[dvipsnames]{xcolor}
\usepackage[normalem]{ulem}
\usepackage{soul}
\soulregister\citet7
\soulregister\citep7

\begin{document}

   \title{A closer look at the young stellar group around Sh~2-295}

   \author{J. V. Corr\^{e}a-Rodrigues\inst{1}
        \and J. Gregorio-Hetem\inst{1}
        }

   \institute{Universidade de S\~{a}o Paulo, Instituto de Astronomia, Geofísica e Ciências Atmosféricas, 05508-090, S\~{a}o Paulo, Brazil\\
             \email{jv.crodrigues27@gmail.com}
            }

   \date{Received September 30, 2025}

  \abstract
    {Star formation is governed by multiple physical processes, making it inherently complicated. One excellent example is the Canis Major OB1/R1 Association, whose complex history of star formation is related to different episodes. Three supernova (SN) events potentially altered the environment and impacted star formation and stellar evolution.}
   {Prior investigations revealed two stellar groups of different ages associated with GU~CMa and Z~CMa. This work focusses on identifying the low-mass young stellar population near FZ~CMa, located between these two groups and spatially related to the H~II region Sh~2-295. Our main goal is to verify whether this group is age-mixed and characterise its physical properties.} 
   {We analysed multi-object spectroscopic data acquired with Gemini/GMOS to search for typical features of T Tauri stars~(TTs) and to determine their spectral types. Lithium absorption line ($\lambda$ 6708~\AA) was used as a youth indicator, while H$\alpha$ emission was investigated to probe accretion activity. We also derived ages based on optical photometry from {\it Gaia} DR3 and compared the projected spatial distribution to diffuse infrared (IR) emission.}
   {We identified 29 TTs, including six new members of the association and three Classical TTs~(CTTs). The equivalent width of the Li~I absorption line suggests an age of $8.1^{+2.1}_{-3.8}$~Myr, while optical photometric data indicate stellar ages ranging from $\sim$1 to 14~Myr. Younger stars are concentrated around Sh~2-295, whereas the older ones are more widely dispersed.}
   {We increased the number of known TTs related to the CMa association. Our results support a scenario of multiple star-formation episodes, including a younger group that may have been triggered by the expansion of Sh~2-295. The influence of SN events appears limited in this context.}


\maketitle

\nolinenumbers

\section{Introduction} \label{sec: intro}

The Canis Major OB1/R1 Association is one of the most intriguing star-forming regions in our Galaxy. It is located approximately 1 kpc away  \citep{gregorio-hetem2008, Santos-Silva2021,Dong2024} and spans a large area \citep[over 80 sq. deg.][]{Fernandes2019}. The region encompasses several reflection and emission nebulae including the H~II regions Sh~2-292, Sh~2-293, Sh~2-295, Sh~2-296, and Sh~2-297 \citep{Sharpless1959}. These nebulae are mainly located in the central part of the Association coinciding with the main molecular clouds (222~deg~$< l < $~227~deg, -3~deg~$< b <$~0~deg). This central region (hereafter CMa region), contains  stellar groups spanning a wide age range \citep[$\sim$1.5 -- 18~Myr ---][]{Fernandes2015,Santos-Silva2021}. Despite extensive studies, the star formation history of CMa remains unclear.

Sh~2-296 is the most notable nebula in CMa. It has an arc-like shape and, together with the absence of luminous stellar objects and the presence of the runaway star HD 54662, led \citet{Herbst1977} to propose that a supernova (SN) triggered star formation in CMa.
Alternatively, \citet{Comeron1998} suggested that the star formation occurred in the pre-existing clouds, and that the stars were subsequently compressed and accelerated by the SN explosion.

More recently, \citet{Fernandes2019} proposed that Sh~2-296 is part of a larger, roughly ellipsoidal structure named CMa-\textit{shell}, with a diameter of about 60~pc. They also identified three runaway stars likely ejected from a similar location within the shell. These features were interpreted as consequences of three successive SN explosions that happened $\sim$ 6~Myr, 2~Myr, and 1~Myr ago. However, the formation of the older stellar population could not be linked to these events. Therefore, the authors suggested that the SNe had only a minimal role in triggering star formation.

In order to reconstruct a three-dimensional morphology and kinematics of gas distribution compared with the stellar population of the CMa region, \citet{Dong2024} used $^{12}$CO, $^{13}$CO, and C$^{18}$O (1-0) along with astrometric data from {\it Gaia} Data Release 3 \citep[DR3,][]{GaiaDR32023}. Their results support the presence of a slowly expanding shell (1.6 $\pm$ 0.7~km~s$^{-1}$), formed by at least two SN events, in agreement with the findings of \citet{Fernandes2019}. 

A 5~square degree X-ray survey conducted using the ROSAT satellite in the western part of the shell revealed the existence of two groups of differing ages, named after the nearest bright stars: GU~CMa and Z~CMa \citep{gregorio-hetem2009}. The GU~CMa group lies further from the main molecular cloud, in contrast to the Z~CMa group. \citet{Santos_Silva2018} revisited part of this area with more sensitive XMM-Newton observations. They found that most stars near Z~CMa are younger than 5~Myr, whereas the bulk of the GU~CMa population is older than 10~Myr. These results support a two-epoch formation scenario: the first episode occurred $\sim$10~Myr ago across the entire region, and a second, more recent episode ($<$5~Myr) took place in areas where there was still molecular gas. 

\citet{Fernandes2015} identified 58 objects --- 41 confirmed T Tauri stars (TTs) + 17 young star candidates --- close to Z~CMa using X-ray detections and optical spectroscopy. Among the TTs, 17\% are Classical T Tauri (CTT), while 83\% are Weak-lined T Tauri (WTT). Based on infrared (IR) excess, they also estimated a low fraction ($\sim$25\%) of disc-bearing stars compared to other young groups with similar ages (1--2~Myr). This fraction may be related to early dissipation due to shock waves from a SN. 

Our main goal is to improve the census of the low-mass young stars towards the region between Z~CMa and GU~CMa (the ``inter-cluster region'') that includes FZ~CMa, a B2~IVn star \citep{Shevchenko1999} associated with the H~II region Sh~2-295. The stellar population in this area is likely to be mixed in age. Our goal is twofold: (i) to characterise this population to assess whether it is indeed mixed, and (ii) to better understand the possible effects of SN on star formation in the region. This study aims to add another piece to the CMa puzzle. The paper is organised as follows. Section \ref{sec: obs} describes Gemini spectroscopic data acquisition and reduction. Section \ref{sec: spec_an} details the spectral analysis. In Sect. \ref{sec: CMD_and_distribution}, we discuss the ages derived from photometry based on the colour-magnitude diagram (CMD), and compare them with dust distribution (mid-infrared emission). Finally, the main findings and conclusions are summarised in Sect. \ref{sec: conc}.

\section{Observations and data reduction} \label{sec: obs}

Optical spectroscopy was performed in March 2017 using the 8~m Gemini South telescope with the Gemini Multi-Object Spectrograph (GMOS). The observations were part of programmes GS-2013A-Q-68 (pre-imaging and mask preparation) and GS-2017A-Q-23 (spectra acquisition). 

The instrumental setup consisted of the R831 grating centred at 6700~\AA~and 6750~\AA, with a 1\arcsec~slit width and 2$\times$2 pixel binning. Two central wavelengths were chosen to prevent spectral gaps between the detectors. According to Gemini grating specifications\footnote{\url{https://www.gemini.edu/instrumentation/gmos/components#Gratings}}, this configuration yields a resolving power of about 2200 at the blaze wavelength ($\lambda$~7570~\AA). Each spectrum covers different wavelength ranges due to the mask configuration and the resulting variation in incidence angle. However, the whole set of spectra covers, at least, wavelengths from 5995~\AA~to 7412~\AA.

\begin{figure*}[]
   \centering
   \includegraphics[width=12cm]{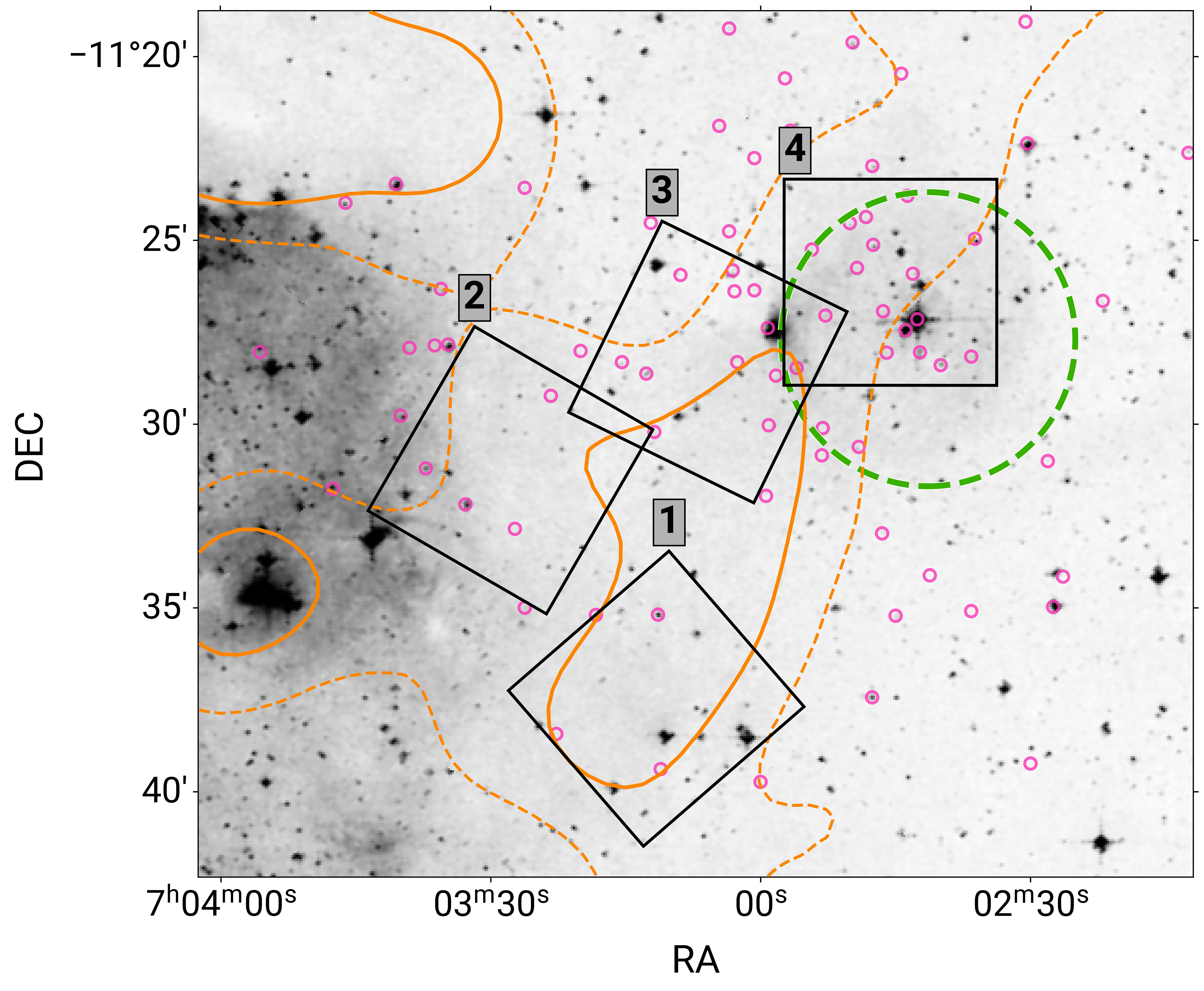}
   \includegraphics[width=14cm]{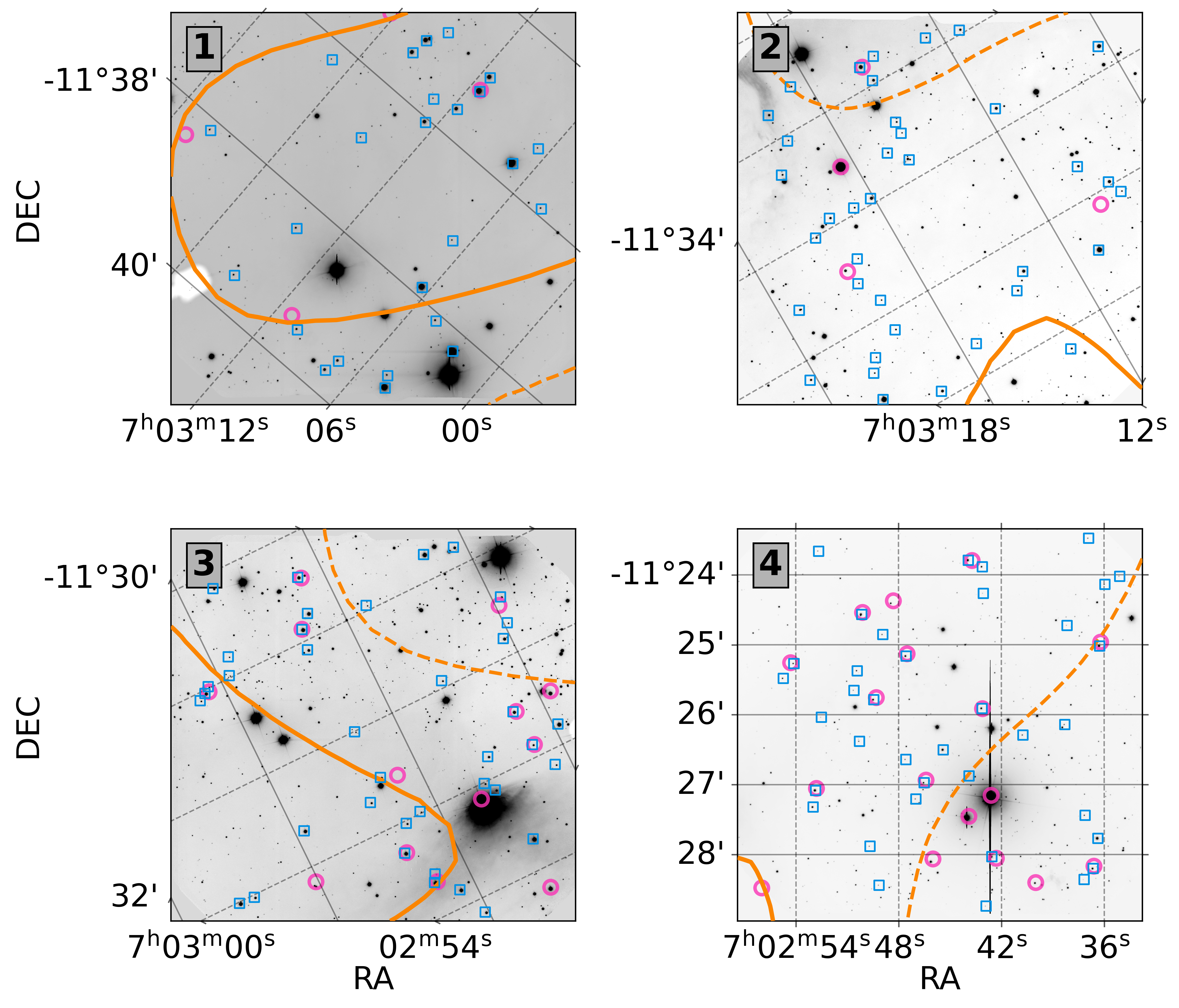}
      \caption{Top panel: optical (Digitized Sky Survey --- DSS2, F+R) image of the region studied in this work. The black squares indicate the GMOS 1-4 fields of Table~\ref{tab: GMOS_fields}. The green dashed circle shows the Sh~2-295 H~II region \citep{Sharpless1959}. 
      Bottom panel: GMOS R pre-images used to prepare GMOS masks. To facilitate the identification of the coordinate system orientation in each image, the grid lines corresponding to RA and DEC are plotted in grey as dashed and solid lines, respectively. The position of the slits are shown by open blue squares. 
      In both panels, orange  lines represent $A_V = 1$ and $A_V = 3$~mag contours (Cambr\'{e}sy, priv. comm.) and open magenta circles indicate X-ray sources \citep{Santos_Silva2018}. Note that not all X-ray sources were covered by the GMOS observations.}
         \label{fig: FZ_CMa_region}
   \end{figure*}

A total of 137 stars were observed across four distinct fields, each with a field-of-view (FoV) of 5.5$^\prime \times 5.5 ^\prime$. The fields were selected to encompass the majority of X-ray sources  previously identified in the ``inter-cluster''~region.
Approximately 18\% of the sample have X-ray counterparts compatible with YSOs \citep{Santos_Silva2018}. The additional candidates were included to complete the observable field. Each target was observed with an exposure time of 1030~s.
Figure~\ref{fig: FZ_CMa_region} shows the GMOS fields and all observed stars in comparison with X-ray sources. A summary of the characteristics of each field is presented in Table~\ref{tab: GMOS_fields}.

GMOS spectroscopic data were reduced using the \texttt{GEMINI/IRAF}\footnote{IRAF is distributed by the National Optical Astronomy Observatory, which is operated by the Association of Universities for Research in Astronomy, Inc., under cooperative agreement with the National Science Foundation} package. We followed a standard procedure, as described in the GMOS data reduction cookbook\footnote{US National Gemini Office 2022, GMOS Data Reduction Cookbook (Version 2.0; Tucson: NSF’s National Optical-Infrared Astronomy Research Laboratory), available online at: \url{https://noirlab.edu/science/programs/csdc/usngo/gmos-cookbook/}. } \citep{Merino2022} and similar to that adopted by \citet{Stanghellini2014}. The procedure consists of eight main steps:
bias subtraction;
flat-fielding;
wavelength calibration;
spectra cutting;
cosmic ray and bad pixel rejection;
spectra extraction;
sky subtraction; and
flux calibration.

In addition to the standard calibration process, a procedure was necessary to refine the astrometry. A crossmatch between GMOS, {\it Gaia} DR3 \citep{GaiaDR32023}, and The Two Micron All Sky Survey \citep[2MASS ---][]{2MASS}, was performed by selecting the nearest source within a maximum separation of 1\arcsec. We found quasi-systematic positional offsets relative to the candidate counterparts that varied (0.5 -- 1.3~\arcsec) across the fields. In order to correct these effects of telescope pointing, a constant shift was added to the equatorial coordinates of each source within a given field ($\Delta_\textrm{RA}$,$\Delta_\textrm{DEC}$ given in Table \ref{tab: GMOS_fields}).

\begin{table}[ht!]
\caption{Characteristics of the observed GMOS fields}                 
\label{tab: GMOS_fields}    
\centering                        
\begin{tabular}{c c c c c c}      
\hline\hline               
 & RA & DEC & N. of & $\Delta_\textrm{RA}$ & $\Delta_\textrm{DEC}$\\    
& (h m s) & (d m s) & targets & (\arcsec) & (\arcsec) \\
\hline                      
1 & 07\,03\,11.69 & -11\,37\,29.6 & 25 & -1.055 & 0.148 \\
2 & 07\,03\,27.73 & -11\,31\,13.8 & 37 & 0.373 & -0.846 \\
3 & 07\,03\,05.81 & -11\,28\,17.3 & 38 & 1.241 & -0.674 \\
4 & 07\,02\,45.71 & -11\,26\,08.8 & 37 & -0.605 & 0.236 \\
\hline                      
\end{tabular}
\tablefoot{Coordinates are given in J2000 reference frame.} 
\end{table}

\section{Spectral analysis} \label{sec: spec_an}

After reducing and calibrating the GMOS spectra, we examined key spectral features typically used to identify low-mass pre-main sequence (PMS) stars. These features allowed us to characterise our sample in terms of accretion activity and stellar youth. In this section, we describe the methodology adopted to analyse the spectroscopic data.

\subsection{Identification of TTs} \label{sec: YSO}

To identify TTs in our sample, we followed the methodology of \citet{Fernandes2015}. The procedure consists of searching for typical spectroscopic features, such as H$\alpha$ emission and absorption in the Li~I line ($\lambda~6708$~\AA). Most of these features were identified in the individual spectra and in the combined spectrum, except in regions affected by artefacts such as cosmic rays.

\begin{figure}[htb!]
   \centering
   \includegraphics[width=\hsize]{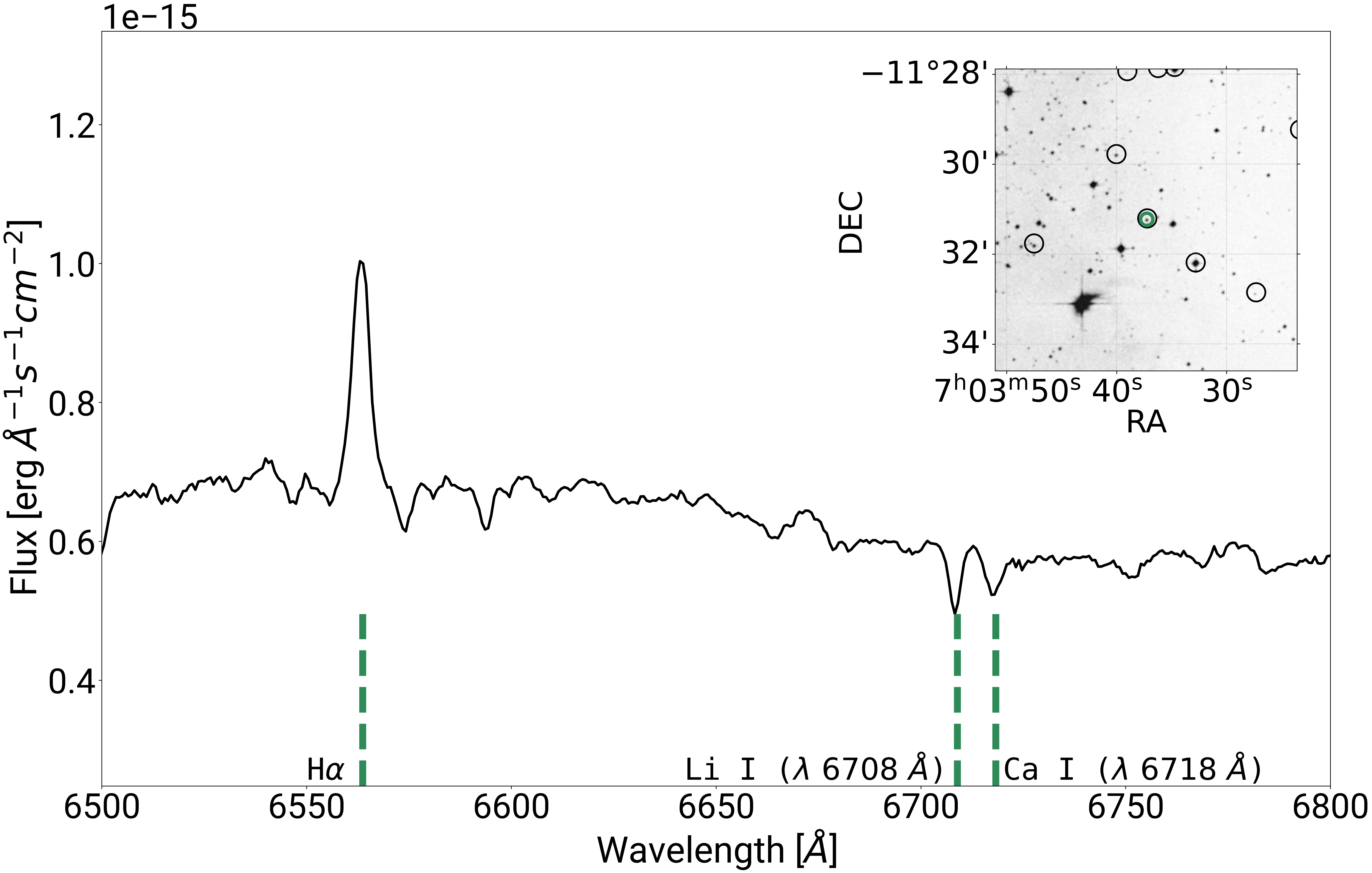}
      \caption{Example of a GMOS spectrum highlighting H$\alpha$ emission, and Li~I ($\lambda~6708$~\AA) and Ca~I ($\lambda$ 6718~\AA) absorption features. The top-right panel shows the angular position of a TTs (green circle) compared to X-ray sources \citep[black circles ---][]{Santos_Silva2018}, confirming it as an X-ray counterpart.}
      \label{fig: TT_id}
   \end{figure}

Among the 137 spectra analysed, we found 24 objects showing both H$\alpha$ emission and Li~I absorption and 5 objects showing only Li~I absorption. These 29 objects were classified as TTs. Table \ref{tab: Yso_list} presents the equivalent widths $W(\textrm{H}\alpha)$ and $W(\textrm{Li})$ of these objects, listed by their 2MASS counterpart identification. The spectrum of the object 2MASS J07033726-1131146 is presented in Fig.~\ref{fig: TT_id} as an example of TTs with H$\alpha$ emission and Li~I absorption. The spectra of the remaining 28 objects are presented in Appendix~\ref{ap: spectra}. We also found seven objects with H$\alpha$ emission but no lithium absorption. Since they are faint stars ($G > 16$ mag) these may be main sequence dwarfs with H$\alpha$ emission (dMe). A list of these objects is provided in Appendix \ref{ap: HalphaEmm}, but they will not be further discussed in this paper.

The signal-to-noise ratio (S/N) of all spectra centred at 6700~\AA~is reported in Tab. \ref{tab: Yso_list} as an indicator of spectral quality. These values were computed using the \texttt{specutils} Python package and represent mean values measured over the 6450 -- 6550~\AA~and 6600 -- 6700~\AA~intervals. The S/N for spectra centred at 6750~\AA~are similar.

To measure the equivalent width ($W$), we followed a procedure similar to that proposed by \citet{Martin1998} and \citet{Alcala2002}. We performed repeated measurements while varying the continuum level. The mean of these measurements was adopted as our $W$ estimate.

Typical relative uncertainties ($\sigma_W/W$) range from 1\% to 10\%, corresponding to $\sigma_W\sim$0.02~\AA~for Li and $\sigma_W\sim$0.1~\AA~for H$\alpha$. These mainly arise from uncertainties in the continuum choice and line profile. Given the medium spectral resolution, unresolved blending with nearby lines --- such as Fe I $\lambda$6710~\AA~--- may have also affected the measurements.

Among the 29 TTs identified in this work, six are new discoveries, while 23 were previously reported in the literature as possible PMS stars. Table~\ref{tab: Yso_list} summarises the available information indicating sources that are X-ray emitters \citep{Santos_Silva2018}; membership probability based on proper motions from {\it Gaia} DR2 \citep{gregorio-hetem2021b}; infrared classification suggesting Class II or Class III objects \citep{Fischer2016,Santos_Silva2018}; and H$\alpha$ emission \citep{Pettersson2019}.

\begin{table*}[htb!]
\caption{List of TTs identified and parameters derived in this work.}
\label{tab: Yso_list}
\centering 
{\footnotesize
\begin{tabular}{l c c c c c c c c c c c } 
\hline\hline 
2MASS & $W$(Li) & $W$(H$\alpha$) & SNR & Sp. T. & $T_{eff}$ & $A_V$ & Class. & Age & CMaX\tablefootmark{a} &  IR Class 
&$P$\tablefootmark{b} \\
& (\AA) & (\AA) &  &  & (K) & (mag) &  & (Myr) &  &   
&\\
\hline

07023628-1125008\tablefootmark{c} & 0.55 & -1.17 & 16 & K7 & 3970 $\pm$ 120 & 1.00 & WTT & 1.00 -- 1.25 & C012 &  $\cdots$ 
&$\cdots$ \\
07023666-1128115 & 0.64 & -3.24 & 15 & K7.5 & 3955 $\pm$ 120 & 1.75 & WTT & $\cdots$ & C019 &  III\tablefootmark{a} 
&$\cdots$ \\
07024312-1123532\tablefootmark{c,d} & 0.57 & -42.46 & 19 & K7 & 3970 $\pm$ 120 & 0.75 & CTT & 5.75 -- 6.00 & C026b &  $\cdots$ 
&66.0 \\
07024314-1125544\tablefootmark{c} & 0.68 & -1.41 & 20 & K7.5 & 3955 $\pm$ 120 & 1.00 & WTT & 2.50 -- 2.75 & C035 &  $\cdots$ 
&88.0 \\
07024393-1123475\tablefootmark{c} & 0.47 & 0.32 & 26 & K7 & 3970 $\pm$ 240 & 0.00 & WTT & 1.50 -- 1.75 & C026a &  ??\tablefootmark{a} 
&95.0 \\
07024651-1126582\tablefootmark{c,e} & 0.62 & -5.44 & 20 & K7.5 & 3955 $\pm$ 120 & 1.25 & WTT & 3.00 -- 3.25 & C023b &  III\tablefootmark{a} 
& $\cdots$ \\
07024757-1125096\tablefootmark{c} & 0.52 & -11.91 & 21 & K7 & 3970 $\pm$ 120 & 1.25 & CTT & 2.25 -- 2.50 & C020a &  ??\tablefootmark{a} 
&87.0 \\
07024758-1126382 & 0.57 & -3.15 & 8 & M2 & 3490 $\pm$ 120 & 0.75 & WTT & 6.75 -- 7.00 & $\cdots$ &  $\cdots$ 
&$\cdots$ \\
07024945-1125470\tablefootmark{d} & 0.60 & -6.32 & 24 & K7.5 & 3955 $\pm$ 120 & 1.75 & WTT? & $\cdots$ & C030 &  III\tablefootmark{a} 
&$\cdots$ \\
07025016-1124341\tablefootmark{d} & 0.51 & -6.85 & 24 & K7 & 3970 $\pm$ 120 & 1.75 & WTT? & $\cdots$ & C003 &  III\tablefootmark{a} 
&$\cdots$ \\
07025284-1127048\tablefootmark{c} & 0.43 & -0.23 & 31 & K2 & 4760 $\pm$ 240 & 1.25 & WTT & 3.00 -- 3.25 & C022 &  III\tablefootmark{a} 
&$\cdots$ \\
07025410-1125160 & 0.40 & -3.50 & 12 & M1.5 & 3560 $\pm$ 120 & 1.00 & WTT & 3.25 -- 3.50 & C044 &  II/III\tablefootmark{a} 
&91.0 \\
07025472-1125288 & 0.47 & -2.07 & 19 & K7.5 & 3955 $\pm$ 120 & 1.75 & WTT & $\cdots$ & $\cdots$ &  $\cdots$ 
&$\cdots$ \\
07025501-1128140\tablefootmark{c} & 0.55 & -0.65 & 19 & K6 & 4020 $\pm$ 120 & 0.25 & WTT & 2.75 -- 3.00 & $\cdots$ &  $\cdots$ 
&83.0 \\
07025603-1128310\tablefootmark{c} & 0.38 & 0.85 & 12 & K2 & 4760 $\pm$ 240 & 2.50 & WTT & 1.75 -- 2.00 & C004a &  $\cdots$ 
&$\cdots$ \\
07025833-1128428\tablefootmark{c,d} & 0.37 & -49.24 & 21 & K0.5 & 4975 $\pm$ 240 & 3.75 & CTT & 4.00 -- 4.25 & C037 &  II\tablefootmark{a,f} 
&$\cdots$ \\
07025934-1127092 & 0.55 & -3.13 & 10 & M2 & 3490 $\pm$ 120 & 0.75 & WTT & 12.25 -- 12.50 & $\cdots$ &  $\cdots$ 
&$\cdots$ \\
07025996-1127154 & 0.67 & -3.52 & 9 & M2 & 3490 $\pm$ 120 & 0.75 & WTT & 9.50 -- 9.75 & $\cdots$ &  $\cdots$ 
&$\cdots$ \\
07030077-1126239 & 0.72 & -2.48 & 16 & K7.5 & 3955 $\pm$ 120 & 2.50 & WTT & 6.75 -- 7.00 & C006 &  III\tablefootmark{a} 
&83.0 \\
07030298-1126263 & 0.46 & -1.88 & 25 & K7.5 & 3955 $\pm$ 120 & 0.75 & WTT & $\cdots$ & C002 &  ??\tablefootmark{a} 
&$\cdots$ \\
07030751-1139396 & 0.52 & -1.56 & 16 & K7.5 & 3955 $\pm$ 120 & 0.50 & WTT & 9.75 -- 10.00 & $\cdots$ &  $\cdots$ 
&$\cdots$ \\
07031139-1135127\tablefootmark{c} & 0.25 & 2.88 & 31 & G3 & 5740 $\pm$ 480 & 1.75 & WTT & 8.00 -- 8.25 & C025 &  III\tablefootmark{a} 
&95.0 \\
07031181-1130169\tablefootmark{c} & 0.40 & -0.61 & 19 & K7 & 3970 $\pm$ 360 & 1.25 & WTT & 4.50 -- 4.75 & C031 &  ??\tablefootmark{a} 
&90.0 \\
07031269-1128382\tablefootmark{c} & 0.41 & 0.48 & 27 & K0.5 & 4975 $\pm$ 240 & 2.50 & WTT & 3.75 -- 4.00 & C029 &  II\tablefootmark{a,f}
&88.0 \\
07031553-1128219\tablefootmark{c} & 0.42 & 0.67 & 27 & K4 & 4330 $\pm$ 120 & 1.75 & WTT & 11.25 -- 11.50 & C097 &  $\cdots$ 
&95.0 \\
07033033-1132028\tablefootmark{c} & 0.65 & -1.98 & 19 & K7.5 & 3955 $\pm$ 120 & 1.50 & WTT & 2.25 -- 2.50 & $\cdots$ &  $\cdots$ 
&$\cdots$ \\
07033123-1129515\tablefootmark{c} & 0.52 & -0.68 & 24 & K7 & 3970 $\pm$ 120 & 1.00 & WTT & 5.25 -- 5.50 & $\cdots$ &  $\cdots$ 
&27.0 \\
07033346-1131113 & 0.53 & -4.93 & 7 & M3.5 & 3260 $\pm$ 120 & 0.75 & WTT & 13.50 -- 13.75 & $\cdots$ &  $\cdots$ 
&$\cdots$ \\
07033726-1131146 & 0.60 & -2.27 & 14 & M0.5 & 3700 $\pm$ 120 & 0.50 & WTT & $\cdots$ & C007 &  $\cdots$ &$\cdots$ \\
\hline
\end{tabular}}
\tablefoot{ Previous results: 
\tablefoottext{a}{\citet{Santos_Silva2018},}
\tablefoottext{b}{\citet{gregorio-hetem2021b},}
\tablefoottext{c}{Members of vdBergh 92 \citep{He2022},}
\tablefoottext{d}{H$\alpha$ emission objects \citep{Pettersson2019},}
\tablefoottext{e}{RS Canum Venaticorum–type systems \citep{Chen2020},}
\tablefoottext{f}{\citet{Fischer2016}.}
}
\end{table*}

\subsection{Spectral classification} \label{sec: spt_type}

Spectral types were estimated  using two different methods: (i) numerical and visual comparison with spectra of known spectral types; and (ii) spectral indices sensitive to spectral type. Both methods are described as follows. 

In the first method, we compared the GMOS spectra with two libraries of young star templates. To perform it consistently, all spectra were degraded to the lower resolution of each pair, and comparisons were made using the \texttt{template\_match} function from the \texttt{specutils} Python package \citep{nicholas_earl_2025_SPECUTILS} to calculate a goodness-of-fit expressed by:
\begin{equation}
    G = \sum_\lambda \left[ \frac{C_{A_V}(\lambda) - \alpha T(\lambda)}{\sigma_C(\lambda)} \right]^2 . 
\end{equation}
$C_{A_V}(\lambda)$ is the observed GMOS spectrum including the effects of visual extinction ($A_V$), $\sigma_C(\lambda)$
are the associated uncertainties, and $T(\lambda)$ is the template spectrum. $\alpha$ is the flux-scaling factor determined by the function \texttt{\_normalize\_for\_template\_matching} from \texttt{specutils} as well. This method is adapted from  \citet{Manjavacas2020}  and \citet{Cushing2008}.

Two distinct libraries of spectral types were used separately: 

\begin{itemize}

\item[\textbullet] a subsample from \citet[][L18]{Luhman_2018} with 14 spectra  (from stars with ages $\sim$11~Myr)  covering K6--M7 types, with a typical uncertainty of 0.25 spectral types. . The resolution varies between 3 and 4~\AA~and the extinction is relatively low ($A_V \lesssim 1.0$~mag). For spectral types earlier than K2, we included spectra of dwarfs (or subgiants when dwarfs were not available) from STELIB \citep{LeBorgne2003}.  These objects have different surface gravities when compared to their younger counterparts and may exhibit significant extinction. Although this library is not specific to young stars, no systematic effect is introduced because our method for this range of earlier spectral types is based on the continuum shape, which is not affected by differences in surface gravity.

\item[\textbullet] a grid of 28  spectra of photospheric templates from C24. This grid is dereddened and covers G5 to M9.5 spectral types, although it is incomplete, especially for G-type stars. 
\end{itemize}

For each object the best-fitting spectral type and $A_V$ were estimated simultaneously by minimising $G$. To do this, the $A_V^{st}$ that best fits each spectral type was first determined by varying it from 0 to 5.75~mag using 0.25 mag steps. This range was adopted according to the limits indicated by dust extinction maps available for the CMa region. Spectra were dereddened using the \texttt{dust\_extinction} python package and the extinction curve from \citet{Gordon2023}. We adopted $R_V = A_V/E(B-V) = 3.1$ as a typical Galactic value, although larger values are possible in dense environments \citep[e.g.][]{Fitzpatrick1999}. The second step was to compare each pair [spectral type, $A_V^{st}$] in order to find the one that best reproduces the observed data. Visual inspection was used to confirm or refine the classification. An example of the spectral type and $A_V$ determined by this method is presented in Fig. \ref{fig: spec_type}. Comparisons between our $A_V$ estimates and other methodologies are presented in Appendix \ref{ap: ext}.

The wavelength intervals used in spectral comparison were selected to account for both temperature-sensitive features and/or the continuum. We considered the regions listed in \citet[][C24]{Claes2024} and those required to calculate the spectral indices from \citet[][HH14]{Herczeg2014} --- some of these regions were also used in the second method. Table~\mbox{\ref{tab: spec_reg}} lists the ranges and corresponding spectral features. Regions affected by telluric bands, strong emission lines (e.g. H$\alpha$), or spectral defects were excluded.

    \begin{table}[ht!]
    \caption{Spectral regions used to spectral classification.}        
    \label{tab: spec_reg}    
    \centering                        
    \begin{tabular}{c c c}      
    \hline\hline               
         $\lambda \lambda$ & Feature & Reference\\  
         (\AA)  & & \\ \hline
         5847 -- 6058&  TiO & C24\\ 
         6080 -- 6390&  TiO & C24\\ 
         6430 -- 6465&  Continuum & HH14 \\ 
         6990 -- 7150&  TiO + CaH & HH14; C24\\ 
         7350 -- 7550&  VO & C24\\
    \hline                             
    \end{tabular}
    \end{table}

   \begin{figure}[htb!]
   \centering
   \includegraphics[width=\hsize]{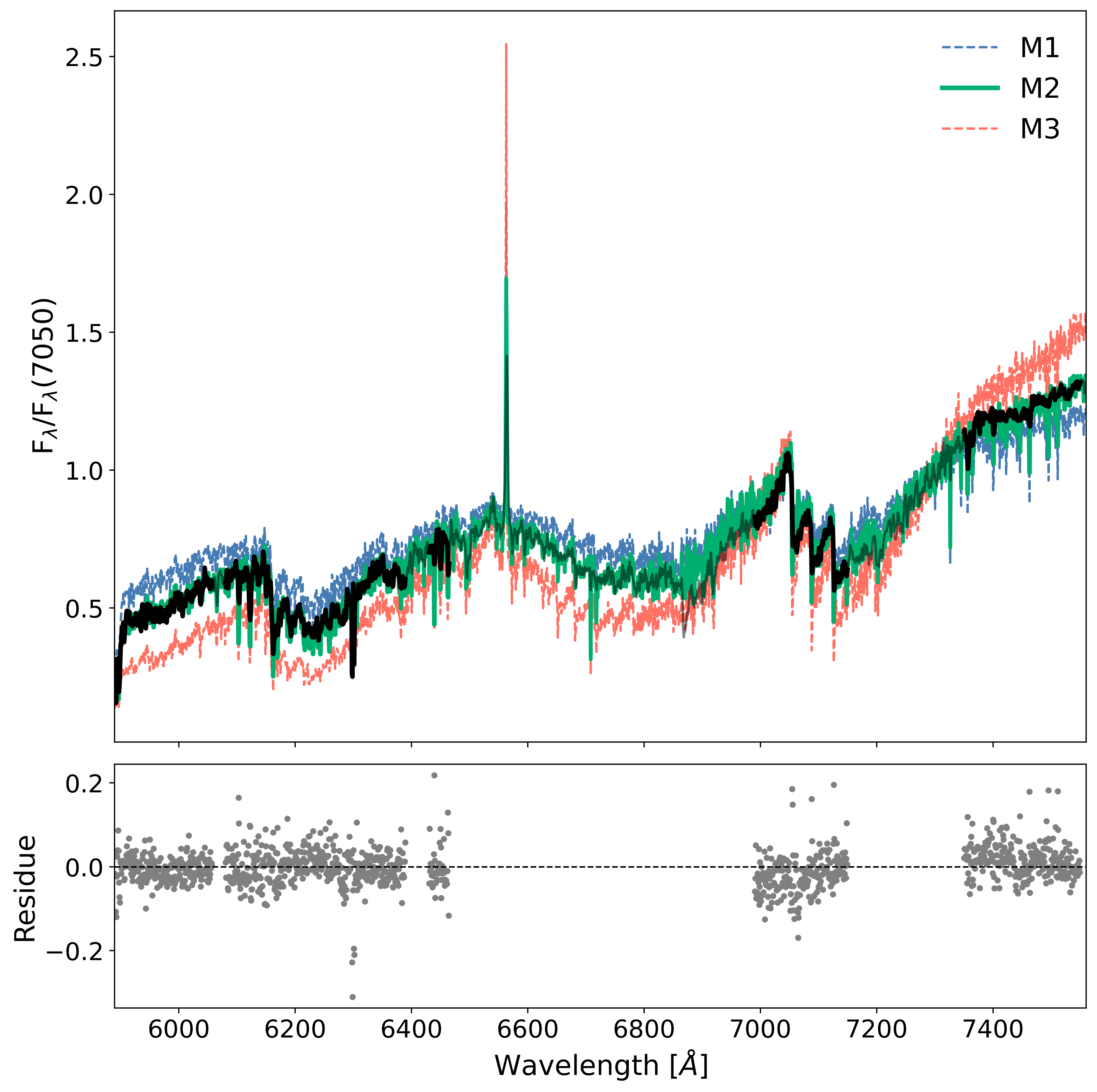}
      \caption{Example of spectral typing via spectral comparison. The GMOS spectrum, normalised by the flux at 7050~\AA, is shown in black in the regions used to calculated $G$ and in grey elsewhere. The pair $A_V = 0.75$~mag, and M2-type stellar spectrum (green line) represents the best fit to the object. For comparison, spectra of M1- and M-3 type stars are shown in blue and orange, respectively. The bottom panel shows the residuals of the comparison between the GMOS object and the M2-type stellar spectrum.}
         \label{fig: spec_type}
   \end{figure}

In the second method, we computed three spectral indices: TiO 6800 (ideal for K5 -- M0.5) and TiO 7140 (for M0 -- M4.5) from HH14, and TiO index from \citet[][J07 --- designed to K5 -- M0 stars]{Jeffries2007}. These indices provide a method of spectral typing by comparing the flux of different spectral narrow  ranges that are sensitive to temperature to continuum ranges. As mentioned by HH14, the conversion between spectral indices to spectral types is sensitive to S/N and to the relative flux calibration. Gravity and metallicity differences can also affect the results.

Typical discrepancies between the two methods were within one spectral subtype (see Appendix \ref{ap: teff_estimates}), which we adopted as the uncertainty. However, the uncertainty could be larger for earlier type stars, where the grids are limited, molecular bands tend to disappear, and the spectral indices used are less discriminating. Some stars also showed degenerate combinations of $A_V$ and spectral type, introducing larger uncertainties. The results were cross-validated by using photometric data to construct a CMD (see Sect. \ref{sec: CMD}). The adopted values are shown in Table \ref{tab: Yso_list}.

Effective temperatures ($T_{eff}$) were assigned as a function of spectral type by adopting the relationship obtained by \citet{Pecaut2013} for stars between 5 and 30~Myr. The mean temperature step between adjacent spectral types is $\sim$120 K, adopted as the uncertainty for one subtype. For stars with larger spectral type uncertainty, the maximum relative error adopted is about $\sim$9\%. 

It is also important to mention that here we adopt a simple single-temperature model for the GMOS spectra. This may introduce biases in the derived stellar parameters of spotted PMS, as shown by \citet{Paolino2025}.

\subsection{CTT or WTT?} \label{sec: CTT}

CTT exhibit spectral features associated with accretion process, including emission line spectrum, presence of forbidden lines, and photospheric continuum excess \citep[e.g.][]{Barrado}.
In particular, H$\alpha$ emission tends to be broad and may present asymmetries. 
The spectrum of WTTs can have H$\alpha$ in emission as well; however, they show smaller equivalent widths when compared to CTTs \citep{Barrado}. In this case, the emission is associated with chromospheric activity \citep[e.g.][]{Fernandes2015}.

Several criteria were proposed to distinguish between CTT and WTT in the literature. Here we adopt two criteria based on the equivalent width of the H$\alpha$ emission line: the saturation criterion proposed by \citet{Barrado} and the empirical criterion to determine non-veiled TTs from \citet{whiteBasri}.

\begin{figure}[htb!]
   \centering
   \includegraphics[width=\hsize]{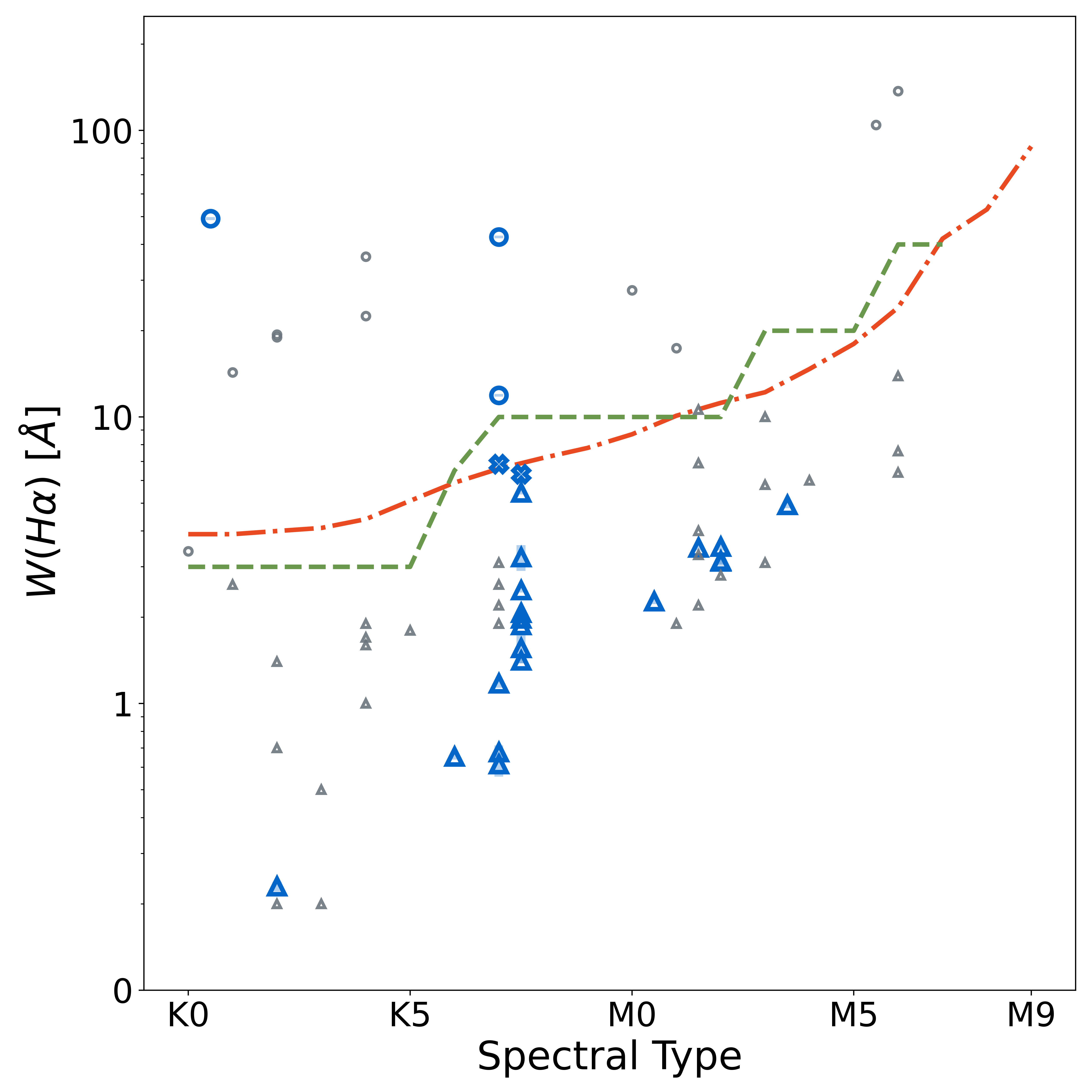}
      \caption{Thresholds defining the CTT classification following \citet[green dashed line][]{whiteBasri} and \citet[orange dash-dotted line][]{Barrado}. Blue symbols represent our sample; grey symbols are data from \citet{Fernandes2015}. These criteria were used to classify the CTTs (indicated by circles), WTTs (triangles), and possible WTTs (crosses).}
      \label{fig: CTT_class}
   \end{figure}

Stars with $W(\textrm{H}\alpha)$ above the thresholds proposed by these authors are classified as CTTs; otherwise, they are considered WTTs. Figure \ref{fig: CTT_class} shows the $W(\textrm{H}\alpha)$ for the 24 stars with H$\alpha$ emission compared to both these thresholds. For clarity, uncertainties in spectral type are not shown, while uncertainties in $W(\textrm{H}\alpha)$ are typically of the same order as the symbol size.

Only three stars meet both criteria and are classified as CTT. One additional star fulfils only the \citet{Barrado} criterion. We consider it as a WTT candidate (WTT?). Considering the adopted uncertainties, the object 07024945-1125470 can also fulfil the \citet{Barrado} criterion. Therefore, it is also considered a WTT candidate. Thus, 10\% of the sample is confirmed as CTTs (excluding WTTs).

For comparison, Fig. \ref{fig: CTT_class} also shows the classification adopted by \citet{Fernandes2015} in a study of the region associated with the Z~CMa system (grey symbols). These authors found a CTT fraction of 17\% in a region where the gas is more concentrated than the region associated with FZ~CMa. Indeed, these fractions should not be considered as representative of the whole region, since a small sample was analysed. 

\subsection{Lithium} \label{sec: Li}

One valuable indicator of youth in the stellar spectrum is the presence of lithium \citep[e.g.][]{Fernandes2015, Pecaut2016, Jeffries2007}. At very young ages, while the central temperatures are lower than $2.5 \times 10^6$~K, lithium is preserved. As the star evolves to the main sequence, the central temperature increases, convective mixing transports Li to the inner regions, and the element is depleted via proton capture \citep{Ushomirsky1998}. Thus, in low-mass stars, a high photospheric lithium abundance is observed only in very young stars \citep{Pecaut2016}. Therefore, this is a distance-independent indicator of youth \citep{Soderblom2014}.

When various clusters in different evolutionary stages are compared, a trend of decreasing lithium abundance  with age is clearly seen, as, for instance, in \citet{Pecaut2016} and \citet{Jeffries2023}. 

\begin{figure}[htb!]
   \centering
   \includegraphics[width=\hsize]{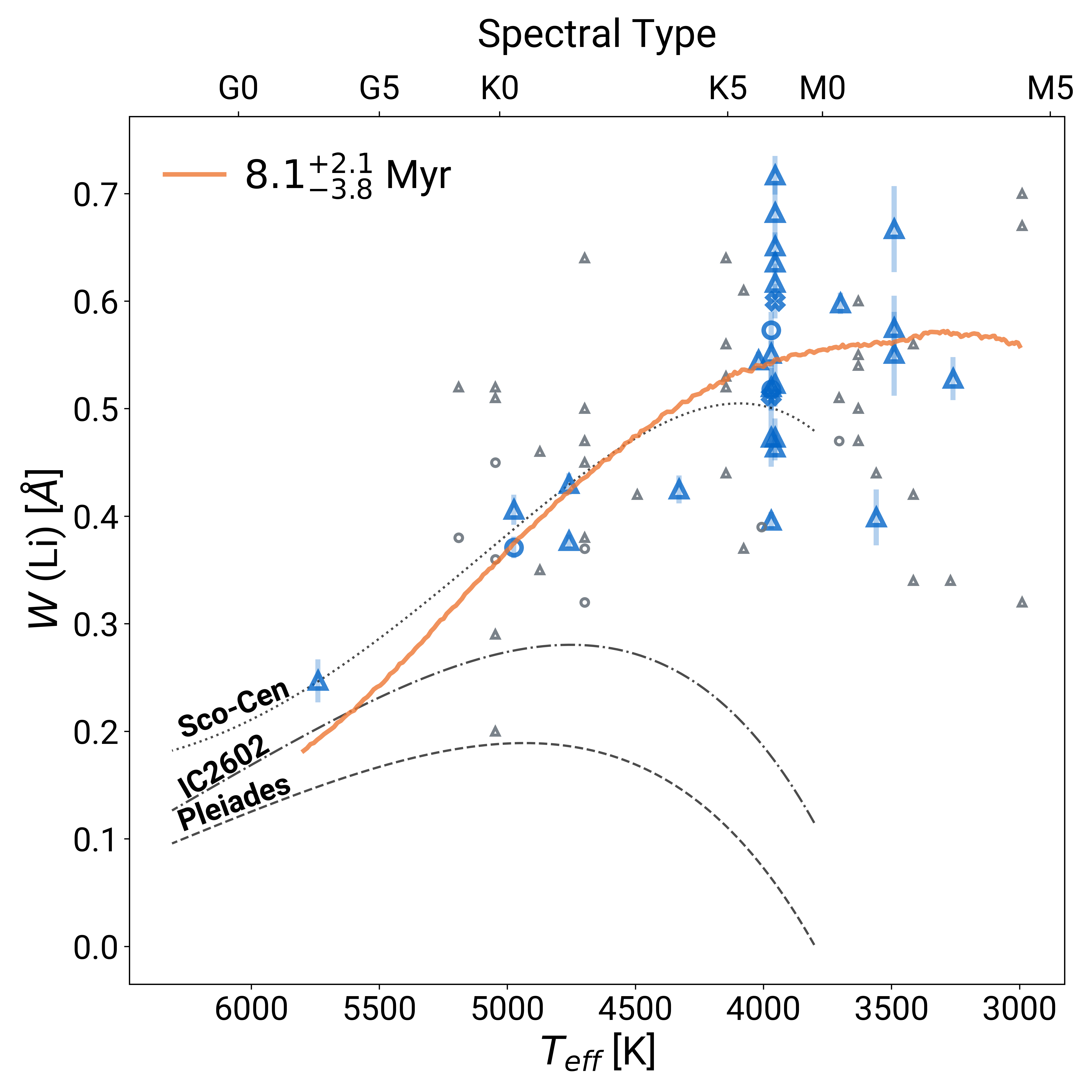}
      \caption{ $W$(Li) from the 6708~\AA~line vs $T_{eff}$ and spectral type. Symbols and colours are the same as in Fig. \ref{fig: CTT_class}. Grey lines are polynomial fits for observed data in other young clusters: Sco-Cen (10–16~Myr), IC2602 (45~Myr) and Pleiades (125~Myr) — see \citet[][]{Pecaut2016} and references therein. The orange line represents the expected curve for a population of $8.1^{+2.1}_{-3.8}$~Myr, as calculated by \texttt{EAGLES v2}.}
      \label{fig: Li}
   \end{figure}

In Fig. \ref{fig: Li}, we show $W(\textrm{Li})$ as a function of effective temperature for the FZ~CMa and Z~CMa groups. A scatter trend is similar for both, which may reflect an age dispersion, different rotation rates, or accretion history, as pointed out by \citet{Fernandes2015}.

Polynomial fits from \citet[][and references therein]{Pecaut2016} for Sco-Cen ($\sim$10-16~Myr), IC 2602 ($\sim$45~Myr), and Pleiades ($\sim$125~Myr) are overplotted for comparison. Our sample lies between Sco-Cen and Z~CMa in this diagram.

We also used the code `Estimating AGes from Lithium Equivalent widthS' (\texttt{EAGLES}) v2 \citep{Jeffries2023,Weaver2024} to infer the most probable age of the FZ~CMa group based on $W(\textrm{Li})$. The code employs an artificial neural network model of the relationship between $T_{eff}$, age, and $W(\textrm{Li})$ also including its intrinsic dispersion. The \texttt{EAGLES v2} code estimates $8.1^{+2.1}_{-3.8}$~Myr for this group, which is in accordance with the age expected, as mentioned before. However, precision is lower for clusters with $\log{\textrm{(age/yr)}} \lesssim 7.1$  when compared to slightly older clusters \citep{Weaver2024}.

\section{Contrasting ages and angular distribution} \label{sec: CMD_and_distribution}

After identifying TTs based on spectroscopic diagnostics, we characterised our sample in terms of age. This allowed us to assess whether the population is indeed mixed, as suggested in previous works. To do that, we used optical photometric data from {\it Gaia} DR3 \citep{GaiaDR32023} and compared it to theoretical models in CMD. This dataset, independent of the GMOS spectra, also served to verify the adopted spectral types.

In Sect. \ref{sec: ang_dist} we explore the angular distribution of the sources to identify potential correlations between spatial location and stellar ages, which could shed light on the formation history of the region.  

\subsection{CMD ages} \label{sec: CMD}

Determining an accurate age estimation for each star could be very challenging, but it is important to disentangle the history of star formation of different clusters.

To constrain the age range of the TTs in our sample, we compared {\it Gaia} DR3 photometric data to PAdova and tRieste Stellar Evolutionary Code (PARSEC) isochrones \citep[release v1.2S + COLIBRI S\_37 + S\_35 + PR16 --- see][and references therein]{Bressan2012,Chen2014, Chen2015,Tang2014, Marigo2017, Pastorelli2019, Pastorelli2020}. Based on previous age estimates for this region, we adopted PARSEC isochrones ranging from 1 to 20~Myr equally spaced by 0.25~Myr.

Among 29 TTs, we found matching photometric data for 23 objects in a 1\arcsec~maximum separation. We only considered sources with re-normalised unit weight error (RUWE) $\leq 1.4$  indicating good  astrometric quality. This criterion was adopted to ensure that reliable counterparts are identified, from which we adopted the photometric data. The remaining six stars did not meet these criteria. To convert apparent magnitudes into absolute magnitudes, we relied on literature values. Two previously reported clusters lie along the same line of sight of the inter-cluster region: VdBerg~92 \citep[e.g.][]{He2022} and CMa~06 \citep[][]{Santos-Silva2021}. \citet{Santos-Silva2021} argue that VdBerg~92 is only part of the CMa~06 group, which was suggested by the spatial distribution and number of members. The astrometric distance of CMa~06 is $1147^{+77}_{-133}$~pc. The inter-cluster region is also located in the direction of subregion B studied by \citet{Dong2024} --- $d = (1087 \pm 19)$~pc. Therefore, we adopted a mean distance of $d = 1117$~pc.

Extinction corrections were applied to the CMD using the relations $A_G/A_V = 0.83627$, $A_{BP}/A_V = 1.08337$, and $A_{RP}/A_V = 0.63439$ \citep{Cardelli1989,Odonnell1994}, respectively corresponding to the effective wavelengths $6390.21$~\AA, $5182.58$~\AA, and $7825.08$~\AA~of the {\it Gaia} filters passbands. These relations are consistent with the extinction curve from \citet{Gordon2023}. The $A_V$ values were adopted from the spectral fitting (Sect.~\ref{sec: spt_type}).

As no degeneracies are present in the CMD region that encompasses our data, we estimated age ranges by finding the closest isochrones to each data point. They were determined by calculating the minimum distance between each source and each isochrone in the colour-magnitude space following the equation: 

\begin{equation}
    d_{min} = \min_i{\sqrt{(x - x_i^{isoc})^2 + (y - y_i^{isoc})^2}} ,
\end{equation}
in which $x$ represents the colour $ (G_{BP}-G_{RP})_0$, $y$ is the absolute magnitude in the {\it Gaia} $G$ band and the subscript $i$ denotes each individual point along the isochrone curve (which has been interpolated to improve accuracy).

\begin{figure}[htb!]
   \centering
   \includegraphics[width=\hsize]{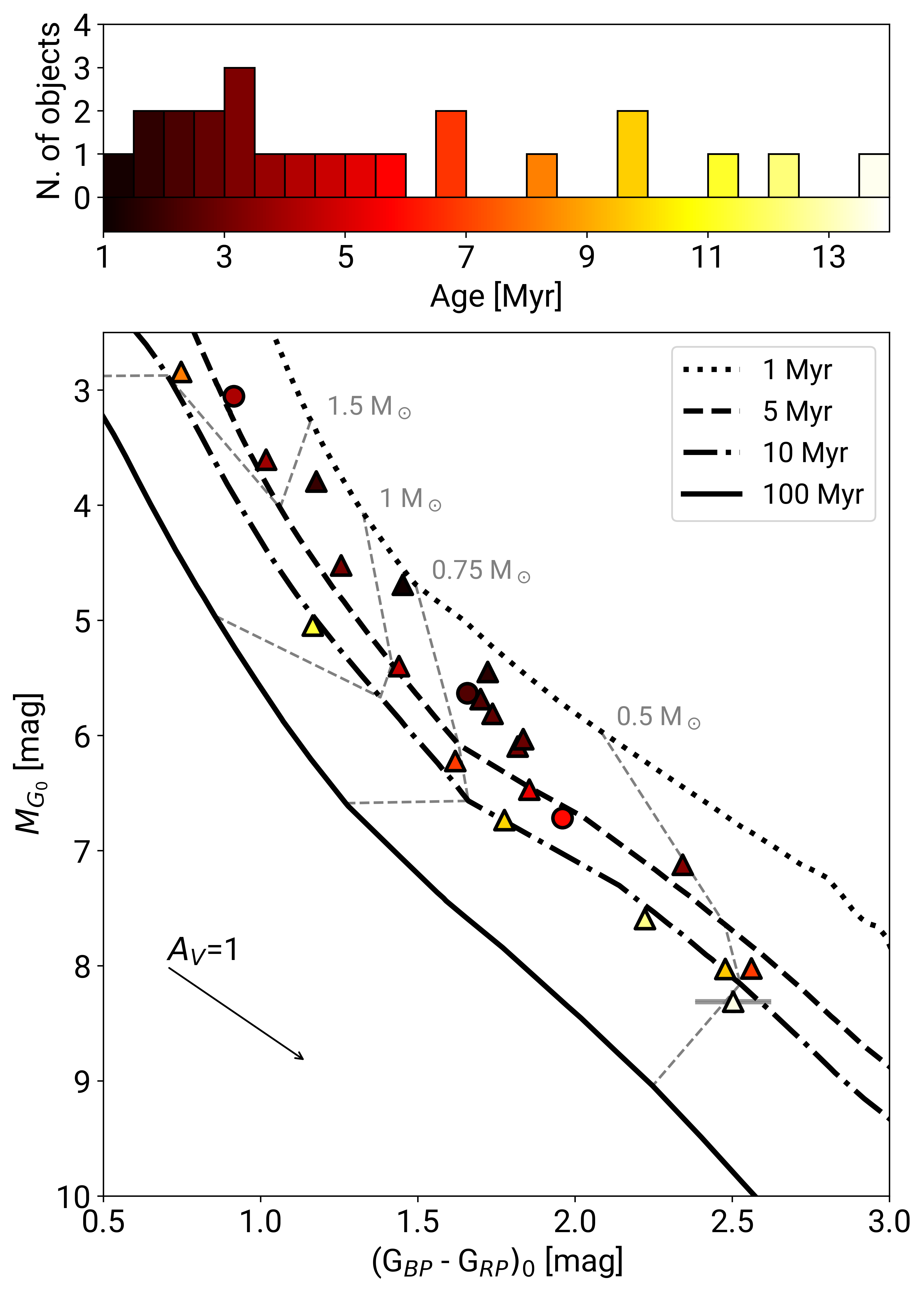}
      \caption{Top panel: histogram of ages estimated for our sample. 
      Bottom: Extinction corrected CMD. The isochrones from PARSEC for 1, 5, 10 and 100~Myr are shown in black thick lines for reference. Grey dotted thin lines connects the points with same mass in each isochrone ranging from 0.5 $M_\odot$ to 1.5 $M_\odot$. 
      In both panels, colour is used to indicate individual ages.}
         \label{fig: CMD}
   \end{figure}

The estimated age ranges are reported in Table~\ref{tab: Yso_list}. We caution that these results depend on multiple uncertainties regarding photometry, extinction estimates, unresolved multiplicity, distance, and model assumptions. The bottom panel of Fig.~\ref{fig: CMD} shows the extinction-corrected CMD. Photometric uncertainties are typically symbol size and tend to be higher for faint sources. An illustration of the expected error bar due to large photometric uncertainties is shown in Fig. \ref{fig: CMD} for the faintest source of the sample.

The top panel of Figure \ref{fig: CMD} shows a histogram of the ages estimated for the FZ~CMa group. The distribution of ages in our sample indicates that 13 stars (57\%) are younger than 5~Myr, 7 (30\%) are 5 -- 10~Myr old, and 3 (13\%) are 10 -- 14~Myr old. It is worth noting that these few older objects are among the faintest, for which the age estimates are less reliable. These results are consistent with what was estimated via lithium depletion, and to what was found for the VdBerg~92 cluster \citep[$\sim$ 5.6~Myr --- ][]{He2022}. It is also consistent with the fact that CMa~06 has a large spread in CMD \citep{Santos-Silva2021}, with most of the stars coinciding with isochrones from 1~Myr to 6~Myr.

\subsection{Angular distribution} \label{sec: ang_dist}

\begin{figure}[htb!]
   \centering
   \includegraphics[width=\hsize]{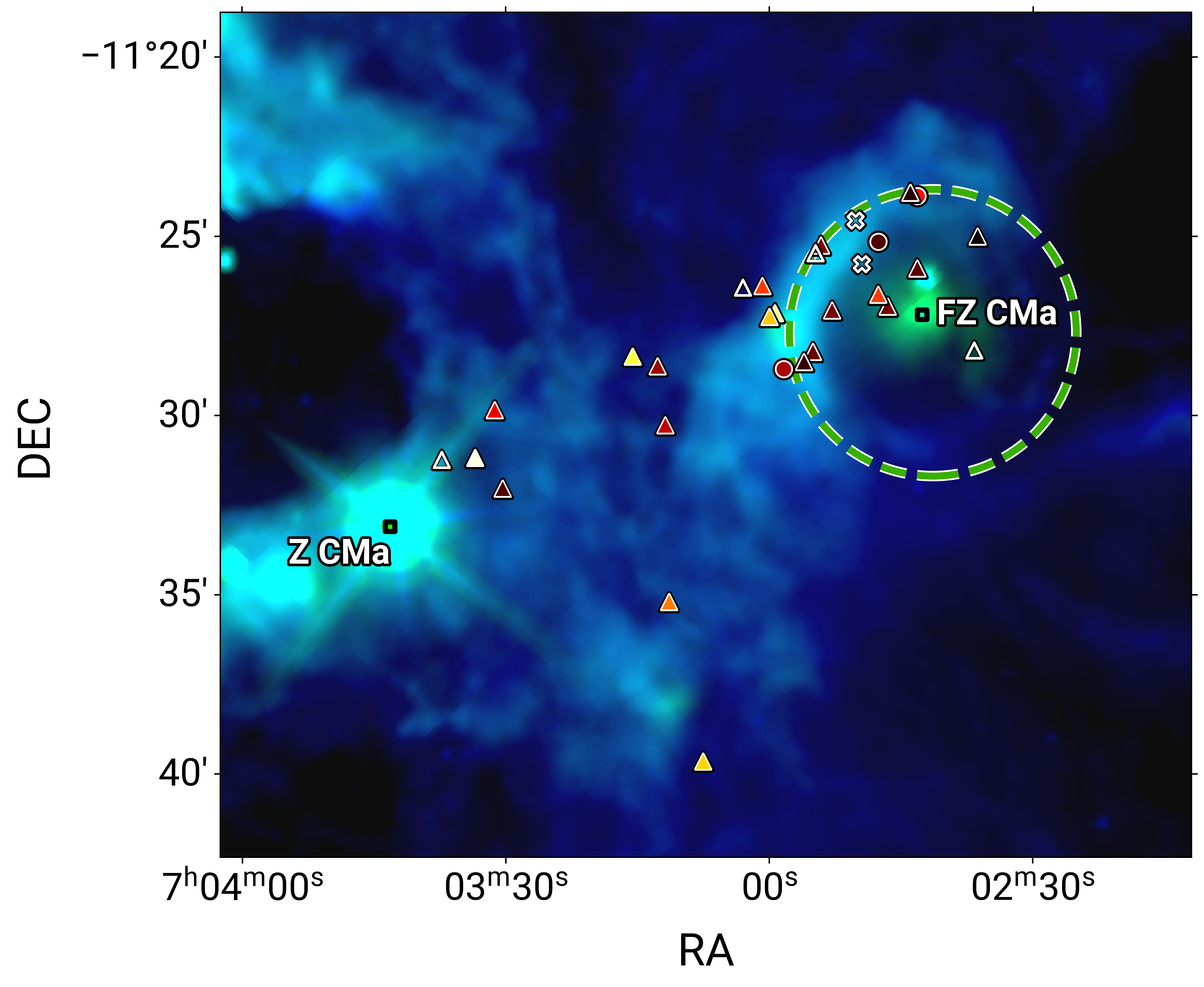}
      \caption{WISE two colour composite: blue is W3, and green is W4. The green dashed circle indicates Sh~2-295 as in Fig. \ref{fig: FZ_CMa_region}. Symbol shapes are the same as in Figs. \ref{fig: CTT_class} and \ref{fig: Li}, while colours follow the age scale from Fig. \ref{fig: CMD}. FZ~CMa and Z~CMa are shown as a reference. The field of view is the same as in the top panel of the Fig. \ref{fig: FZ_CMa_region}.}
         \label{fig: W3band}
   \end{figure}

Figure \ref{fig: W3band} shows the projected spatial distribution of the TTs found in this work overlaid on a two-colour composite with mid-infrared widefield infrared survey explorer \citep[WISE ---][]{WISE} W3 (12 $\mu$m) and W4 (22 $\mu$m) bands. The Sh~2-295 H~II region is clearly seen as an arc-like structure. According to \citet[][and references therein]{Anderson2014}, polycyclic aromatic hydrocarbons (PAHs) molecules are the ones that mainly contribute to emission in W3 band in H~II regions while W4 band traces small dust grains that were stochastically heated.

A spatial correlation is apparent between the position of many TTs and the regions of enhanced IR emission, particularly around Sh~2-295. About half of our sample lies close to Sh~2-295, and this subset appears preferentially located in areas of stronger W3 emission, giving an idea of the PAH distribution. These results are similar to what was found in Sh~2-297 on the distribution of Class II objects and H$\alpha$ emission line stars \citep{Mallick2012}.

To investigate possible subgroups, we colour-coded the symbols by CMD-derived age range (Sect. \ref{sec: CMD}). Younger stars in our sample ($<$5~Myr) appear more concentrated around Sh~2-295: 10 out of 15 stars that are close to Sh~2-295 are $\lesssim 5$~Myr, whereas only 2 of the very young stars are in the other portions of the region (considering only stars that have {\it Gaia} data available).

These results agree with the conclusions of \citet{Santos_Silva2018}. The relatively old population (including FZ~CMa) likely formed in an earlier star formation episode unrelated to the more recent SNe. The SNe may have contributed to triggering star formation (at least of the younger population), but their influence appears limited, as stated by \citet{Fernandes2019}. The projected distribution of the younger TTs around Sh~2-295 may indicate a local episode of star formation, possibly caused by the fragmentation of material around the H~II region.

\section{Summary and Conclusions} \label{sec: conc}

In this study, we investigated the low-mass stellar population associated with the FZ~CMa system. This region lies between two concentrations with different ages: one group with stars typically older than 10~Myr and another younger than 5~Myr. Therefore, this region was previously suggested to have a mixed-age population.

To verify this hypothesis, our team acquired multi-object spectroscopic data from more than a hundred objects using the Gemini-South Telescope. By identifying H$\alpha$ emission and Li~I absorption ($\lambda$ 6708~\AA) lines, we found 29 TTs, 20 of which are X-ray counterparts. We confirmed the youth of 23 objects previously studied in the literature and discovered 6 TTs that are new members.

We employed two different methods to determine spectral types: (i) comparison with template spectra, accounting for extinction, and (ii) spectral index measurements. By comparing the H$\alpha$ emission with criteria proposed by \citet{whiteBasri} and \citet{Barrado}, we found three CTT and 18 WTT. This yields a fraction $n_{\textrm{CTT}}/n_{\textrm{WTT}} = $10\%. Previous results from the literature indicate only two Class II objects, one Class II/III, and seven Class III. No WISE classification is available for the remaining sources.

The youth in our sample was indicated by the presence of Li~I ($\lambda$ 6708~\AA)  in the spectra. A mean age of $8.1^{+2.1}_{-3.8}$~Myr was estimated for the whole group using an artificial neural network model. It correlates Li~I equivalent width with the effective temperature of the stars to further infer the group's age. On the other hand, a CMD compared {\it Gaia} DR3 photometric data with theoretical models from PARSEC. It revealed a spread in ages, with stars mainly distributed from $\sim$1 to 14~Myr.

Are these ages and age spreads realistic? As discussed by \citet{Pecaut2016}, evolutionary models often underperform in reproducing parameters for M-type stars, for instance. Moreover, as discussed by \citet{Soderblom2014}, for stars $\lesssim 20$~Myr, model-dependent and observational uncertainties increase significantly, making it very difficult to estimate reliable absolute ages. In this context, the ages derived here should be regarded as indicators rather than precise absolute values.

Comparing the FZ~CMa’s group with clusters of similar age, we found a CTT fraction consistent with what would be expected for the fraction of accreting stars \citep{Fedele2010,Delfini2025}. This is the opposite of what is observed in the region associated with Z~CMa, where \citet{Fernandes2015} found a lower than expected CTT fraction of 17-24\% despite a mean age of 1--2~Myr. However, our sample is very limited, and additional data are needed to draw reliable conclusions about it. 

The projected spatial distribution of TTs, combined with age estimates, supports a scenario where the SN events played a minor role in triggering star formation near FZ~CMa. Instead, we identified an older, spatially dispersed population and a younger group concentrated near Sh~2-295. The latter may have formed due to localised triggering by the expanding H~II region Sh~2-295.

\begin{acknowledgements}
      We acknowledge the São Paulo Research Foundation (FAPESP) for the 
      financial support (grants \#2022/09374-0, \#2023/08726-2).

      We also thank Rodrigo Carrasco for helping with data reduction of the GMOS spectra.

      Based on observations obtained at the international Gemini Observatory, a program of NSF NOIRLab, which is managed by the Association of Universities for Research in Astronomy (AURA) under a cooperative agreement with the U.S. National Science Foundation on behalf of the Gemini Observatory partnership: the U.S. National Science Foundation (United States), National Research Council (Canada), Agencia Nacional de Investigaci\'{o}n y Desarrollo (Chile), Ministerio de Ciencia, Tecnolog\'{i}a e Innovaci\'{o}n (Argentina), Minist\'{e}rio da Ci\^{e}ncia, Tecnologia, Inova\c{c}\~{o}es e Comunica\c{c}\~{o}es (Brazil), and Korea Astronomy and Space Science Institute (Republic of Korea).

      This publication makes use of data products from the Two Micron All Sky Survey, which is a joint project of the University of Massachusetts and the Infrared Processing and Analysis Center/California Institute of Technology, funded by the National Aeronautics and Space Administration and the National Science Foundation.

      This publication makes use of data products from the Wide-field Infrared Survey Explorer, which is a joint project of the University of California, Los Angeles, and the Jet Propulsion Laboratory/California Institute of Technology, funded by the National Aeronautics and Space Administration.

      The Digitized Sky Survey was produced at the Space Telescope Science Institute under U.S. Government grant NAG W–2166. The images of these surveys are based on photographic data obtained using the Oschin Schmidt Telescope on Palomar Mountain and the UK Schmidt Telescope. The plates were processed into the present compressed digital form with the permission of these institutions.

      This work has made use of data from the European Space Agency (ESA) mission {\it Gaia} (\url{https://www.cosmos.esa.int/gaia}), processed by the {\it Gaia} Data Processing and Analysis Consortium (DPAC, \url{https://www.cosmos.esa.int/web/gaia/dpac/consortium}). Funding for the DPAC has been provided by national institutions, in particular the institutions participating in the {\it Gaia} Multilateral Agreement.

      This work has made use of the VizieR catalogue access tool, Aladin sky atlas, and SIMBAD database operated at CDS, Strasbourg Astronomical Observatory, France.

      \textit{Software:} \texttt{Dustmaps} \citep{Green2018_dustmaps}, \texttt{IRAF} \citep{Tody1986_IRAF},
      \texttt{Numpy} \citep{harris2020_numpy},
      \texttt{Pandas} \citep{mckinney-proc-scipy-2010_PANDAS},
      \texttt{Matplotlib} \citep{Hunter2007_Matplotlib},
      \texttt{Scipy} \citep{2020SciPy-NMeth},
      \texttt{Astropy} \citep{2022Astropy1,2018Astropy2,2013Astropy3},
      \texttt{dust\_extinction} \citep{2024Gordon_dust_extinction},
      \texttt{TOPCAT} \citep{Taylor2005_TOPCAT},
      \texttt{Stilts} \citep{Taylor2006_STILTS}.

\end{acknowledgements}

\bibliographystyle{aa} 
\bibliography{references} 

@ARTICLE{Fernandes2015,
       author = {{Fernandes}, B. and {Gregorio-Hetem}, J. and {Montmerle}, T. and {Rojas}, G.},
        title = "{Spectroscopic characterization of X-ray emitting young stars associated with the Sh 2-296 nebula}",
      journal = {MNRAS},
         year = 2015,
        month = mar,
       volume = {448},
       number = {1},
        pages = {119-134},
          doi = {10.1093/mnras/stv001},
archivePrefix = {arXiv},
       eprint = {1501.03763},
 primaryClass = {astro-ph.SR},
       adsurl = {https://ui.adsabs.harvard.edu/abs/2015MNRAS.448..119F}
}

@ARTICLE{whiteBasri,
       author = {{White}, Russel J. and {Basri}, Gibor},
        title = "{Very Low Mass Stars and Brown Dwarfs in Taurus-Auriga}",
      journal = {ApJ},
         year = 2003,
        month = jan,
       volume = {582},
       number = {2},
        pages = {1109-1122},
          doi = {10.1086/344673},
archivePrefix = {arXiv},
       eprint = {astro-ph/0209164},
 primaryClass = {astro-ph},
       adsurl = {https://ui.adsabs.harvard.edu/abs/2003ApJ...582.1109W}
}

@ARTICLE{Barrado,
       author = {{Barrado y Navascu{\'e}s}, David and {Mart{\'\i}n}, Eduardo L.},
        title = "{An Empirical Criterion to Classify T Tauri Stars and Substellar Analogs Using Low-Resolution Optical Spectroscopy}",
      journal = {AJ},
         year = 2003,
        month = dec,
       volume = {126},
       number = {6},
        pages = {2997-3006},
          doi = {10.1086/379673},
archivePrefix = {arXiv},
       eprint = {astro-ph/0309284},
 primaryClass = {astro-ph},
       adsurl = {https://ui.adsabs.harvard.edu/abs/2003AJ....126.2997B}
}

@ARTICLE{2MASS,
       author = {{Skrutskie}, M.~F. and {Cutri}, R.~M. and {Stiening}, R. and {Weinberg}, M.~D. and {Schneider}, S. and {Carpenter}, J.~M. and {Beichman}, C. and {Capps}, R. and {Chester}, T. and {Elias}, J. and {Huchra}, J. and {Liebert}, J. and {Lonsdale}, C. and {Monet}, D.~G. and {Price}, S. and {Seitzer}, P. and {Jarrett}, T. and {Kirkpatrick}, J.~D. and {Gizis}, J.~E. and {Howard}, E. and {Evans}, T. and {Fowler}, J. and {Fullmer}, L. and {Hurt}, R. and {Light}, R. and {Kopan}, E.~L. and {Marsh}, K.~A. and {McCallon}, H.~L. and {Tam}, R. and {Van Dyk}, S. and {Wheelock}, S.},
        title = "{The Two Micron All Sky Survey (2MASS)}",
      journal = {AJ},
         year = 2006,
        month = feb,
       volume = {131},
       number = {2},
        pages = {1163-1183},
          doi = {10.1086/498708},
       adsurl = {https://ui.adsabs.harvard.edu/abs/2006AJ....131.1163S}
}

@ARTICLE{WISE,
       author = {{Wright}, Edward L. and {Eisenhardt}, Peter R.~M. and {Mainzer}, Amy K. and {Ressler}, Michael E. and {Cutri}, Roc M. and {Jarrett}, Thomas and {Kirkpatrick}, J. Davy and {Padgett}, Deborah and {McMillan}, Robert S. and {Skrutskie}, Michael and {Stanford}, S.~A. and {Cohen}, Martin and {Walker}, Russell G. and {Mather}, John C. and {Leisawitz}, David and {Gautier}, Thomas N., III and {McLean}, Ian and {Benford}, Dominic and {Lonsdale}, Carol J. and {Blain}, Andrew and {Mendez}, Bryan and {Irace}, William R. and {Duval}, Valerie and {Liu}, Fengchuan and {Royer}, Don and {Heinrichsen}, Ingolf and {Howard}, Joan and {Shannon}, Mark and {Kendall}, Martha and {Walsh}, Amy L. and {Larsen}, Mark and {Cardon}, Joel G. and {Schick}, Scott and {Schwalm}, Mark and {Abid}, Mohamed and {Fabinsky}, Beth and {Naes}, Larry and {Tsai}, Chao-Wei},
        title = "{The Wide-field Infrared Survey Explorer (WISE): Mission Description and Initial On-orbit Performance}",
      journal = {AJ},
         year = 2010,
        month = dec,
       volume = {140},
       number = {6},
        pages = {1868-1881},
          doi = {10.1088/0004-6256/140/6/1868},
archivePrefix = {arXiv},
       eprint = {1008.0031},
 primaryClass = {astro-ph.IM},
       adsurl = {https://ui.adsabs.harvard.edu/abs/2010AJ....140.1868W}
}

@article{herbst1977,
       author = {{Herbst}, W. and {Assousa}, G.~E.},
        title = "{Observational evidence for supernova-induced star formation: Canis Major R1.}",
      journal = {ApJ},
         year = 1977,
        month = oct,
       volume = {217},
        pages = {473-487},
          doi = {10.1086/155596},
       adsurl = {https://ui.adsabs.harvard.edu/abs/1977ApJ...217..473H}
}

@INCOLLECTION{gregorio-hetem2008,
       author = {{Gregorio-Hetem}, J.},
        title = "{The Canis Major Star Forming Region}",
    booktitle = {Handbook of Star Forming Regions, Volume II},
         year = 2008,
       editor = {{Reipurth}, B.},
       volume = {5},
        pages = {1},
          doi = {10.48550/arXiv.0808.3812},
       adsurl = {https://ui.adsabs.harvard.edu/abs/2008hsf2.book....1G}
}

@article{Luhman_2018,
       author = {{Luhman}, K.~L. and {Herrmann}, K.~A. and {Mamajek}, E.~E. and {Esplin}, T.~L. and {Pecaut}, M.~J.},
        title = "{New Young Stars and Brown Dwarfs in the Upper Scorpius Association}",
      journal = {AJ},
         year = 2018,
        month = aug,
       volume = {156},
       number = {2},
          eid = {76},
        pages = {76},
          doi = {10.3847/1538-3881/aacc6d},
archivePrefix = {arXiv},
       eprint = {1807.07955},
 primaryClass = {astro-ph.SR},
       adsurl = {https://ui.adsabs.harvard.edu/abs/2018AJ....156...76L}
}

@ARTICLE{Cushing2008,
       author = {{Cushing}, Michael C. and {Marley}, Mark S. and {Saumon}, D. and {Kelly}, Brandon C. and {Vacca}, William D. and {Rayner}, John T. and {Freedman}, Richard S. and {Lodders}, Katharina and {Roellig}, Thomas L.},
        title = "{Atmospheric Parameters of Field L and T Dwarfs}",
      journal = {ApJ},
         year = 2008,
        month = may,
       volume = {678},
       number = {2},
        pages = {1372-1395},
          doi = {10.1086/526489},
archivePrefix = {arXiv},
       eprint = {0711.0801},
 primaryClass = {astro-ph},
       adsurl = {https://ui.adsabs.harvard.edu/abs/2008ApJ...678.1372C}
}

@ARTICLE{Manjavacas2020,
       author = {{Manjavacas}, E. and {Lodieu}, N. and {B{\'e}jar}, V.~J.~S. and {Zapatero-Osorio}, M.~R. and {Boudreault}, S. and {Bonnefoy}, M.},
        title = "{Spectral library of age-benchmark low-mass stars and brown dwmarfs}",
      journal = {MNRAS},
         year = 2020,
        month = feb,
       volume = {491},
       number = {4},
        pages = {5925-5950},
          doi = {10.1093/mnras/stz3441},
archivePrefix = {arXiv},
       eprint = {1912.02806},
 primaryClass = {astro-ph.SR},
       adsurl = {https://ui.adsabs.harvard.edu/abs/2020MNRAS.491.5925M}
}

@ARTICLE{Santos_Silva2018,
       author = {{Santos-Silva}, T. and {Gregorio-Hetem}, J. and {Montmerle}, T. and {Fernandes}, B. and {Stelzer}, B.},
        title = "{Star formation history of Canis Major OB1. II. A bimodal X-ray population revealed by XMM-Newton}",
      journal = {A\&A},
         year = 2018,
        month = feb,
       volume = {609},
          eid = {A127},
        pages = {A127},
          doi = {10.1051/0004-6361/201730815},
archivePrefix = {arXiv},
       eprint = {1710.01876},
 primaryClass = {astro-ph.GA},
       adsurl = {https://ui.adsabs.harvard.edu/abs/2018A&A...609A.127S}
}

@ARTICLE{gregorio-hetem2009,
       author = {{Gregorio-Hetem}, J. and {Montmerle}, T. and {Rodrigues}, C.~V. and {Marciotto}, E. and {Preibisch}, T. and {Zinnecker}, H.},
        title = "{Star formation history of Canis Major R1. I. Wide-Field X-ray study of the young stellar population}",
      journal = {A\&A},
         year = 2009,
        month = nov,
       volume = {506},
       number = {2},
        pages = {711-727},
          doi = {10.1051/0004-6361/200912140},
archivePrefix = {arXiv},
       eprint = {0909.2888},
 primaryClass = {astro-ph.SR},
       adsurl = {https://ui.adsabs.harvard.edu/abs/2009A&A...506..711G}
}

@ARTICLE{Stanghellini2014,
       author = {{Stanghellini}, Letizia and {Magrini}, Laura and {Casasola}, Viviana and {Villaver}, Eva},
        title = "{The radial metallicity gradient and the history of elemental enrichment in M 81 through emission-line probes}",
      journal = {A\&A},
         year = 2014,
        month = jul,
       volume = {567},
          eid = {A88},
        pages = {A88},
          doi = {10.1051/0004-6361/201423423},
archivePrefix = {arXiv},
       eprint = {1403.5547},
 primaryClass = {astro-ph.GA},
       adsurl = {https://ui.adsabs.harvard.edu/abs/2014A&A...567A..88S}
}

@ARTICLE{gregorio-hetem2021b,
       author = {{Gregorio-Hetem}, J. and {Navarete}, F. and {Hetem}, A. and {Santos-Silva}, T. and {Galli}, P.~A.~B. and {Fernandes}, B. and {Montmerle}, T. and {Jatenco-Pereira}, V. and {Rorges Fernandes}, M. and {Perottoni}, H.~D. and {Schoenell}, W. and {Ribeiro}, T. and {Kanaan}, A.},
        title = "{Searching for Active Low-mass Stars in the CMa Star-forming Region: Multi-band Photometry with T80S}",
      journal = {AJ},
         year = 2021,
        month = mar,
       volume = {161},
       number = {3},
          eid = {133},
        pages = {133},
          doi = {10.3847/1538-3881/abd705},
archivePrefix = {arXiv},
       eprint = {2012.15166},
 primaryClass = {astro-ph.SR},
       adsurl = {https://ui.adsabs.harvard.edu/abs/2021AJ....161..133G}
}

@ARTICLE{Santos-Silva2021,
       author = {{Santos-Silva}, T. and {Perottoni}, H.~D. and {Almeida-Fernandes}, F. and {Gregorio-Hetem}, J. and {Jatenco-Pereira}, V. and {Mendes de Oliveira}, C. and {Montmerle}, T. and {Bica}, E. and {Bonatto}, C. and {Monteiro}, H. and {Dias}, W.~S. and {Barbosa}, C.~E. and {Fernandes}, B. and {Galli}, P.~A.~B. and {Borges Fernandes}, M. and {Kanaan}, A. and {Ribeiro}, T. and {Schoenell}, W.},
        title = "{Canis Major OB1 stellar group contents revealed by Gaia}",
      journal = {MNRAS},
         year = 2021,
        month = nov,
       volume = {508},
       number = {1},
        pages = {1033-1055},
          doi = {10.1093/mnras/stab2409},
archivePrefix = {arXiv},
       eprint = {2108.06234},
 primaryClass = {astro-ph.GA},
       adsurl = {https://ui.adsabs.harvard.edu/abs/2021MNRAS.508.1033S}
}

@ARTICLE{Martin1998,
       author = {{Martin}, E.~L. and {Montmerle}, T. and {Gregorio-Hetem}, J. and {Casanova}, S.},
        title = "{Spectroscopic classification of X-ray selected stars in the rho Ophiuchi star-forming region and vicinity}",
      journal = {MNRAS},
         year = 1998,
        month = nov,
       volume = {300},
       number = {3},
        pages = {733-746},
          doi = {10.1046/j.1365-8711.1998.01932.x},
       adsurl = {https://ui.adsabs.harvard.edu/abs/1998MNRAS.300..733M}
}

@ARTICLE{Mallick2012,
       author = {{Mallick}, K.~K. and {Ojha}, D.~K. and {Samal}, M.~R. and {Pandey}, A.~K. and {Bhatt}, B.~C. and {Ghosh}, S.~K. and {Dewangan}, L.~K. and {Tamura}, M.},
        title = "{Star Formation Activity in the Galactic H II Region Sh2-297}",
      journal = {ApJ},
         year = 2012,
        month = nov,
       volume = {759},
       number = {1},
          eid = {48},
        pages = {48},
          doi = {10.1088/0004-637X/759/1/48},
archivePrefix = {arXiv},
       eprint = {1209.3420},
 primaryClass = {astro-ph.GA},
       adsurl = {https://ui.adsabs.harvard.edu/abs/2012ApJ...759...48M}
}

@ARTICLE{Pettersson2019,
       author = {{Pettersson}, Bertil and {Reipurth}, Bo},
        title = "{H{\ensuremath{\alpha}} emission-line stars in molecular clouds. III. Canis Major}",
      journal = {A\&A},
         year = 2019,
        month = oct,
       volume = {630},
          eid = {A90},
        pages = {A90},
          doi = {10.1051/0004-6361/201731578},
archivePrefix = {arXiv},
       eprint = {1908.11446},
 primaryClass = {astro-ph.SR},
       adsurl = {https://ui.adsabs.harvard.edu/abs/2019A&A...630A..90P}
}

@ARTICLE{Fedele2010,
       author = {{Fedele}, D. and {van den Ancker}, M.~E. and {Henning}, Th. and {Jayawardhana}, R. and {Oliveira}, J.~M.},
        title = "{Timescale of mass accretion in pre-main-sequence stars}",
      journal = {A\&A},
         year = 2010,
        month = feb,
       volume = {510},
          eid = {A72},
        pages = {A72},
          doi = {10.1051/0004-6361/200912810},
archivePrefix = {arXiv},
       eprint = {0911.3320},
 primaryClass = {astro-ph.SR},
       adsurl = {https://ui.adsabs.harvard.edu/abs/2010A&A...510A..72F}
}

@ARTICLE{Shevchenko1999,
       author = {{Shevchenko}, V.~S. and {Ezhkova}, O.~V. and {Ibrahimov}, M.~A. and {van den Ancker}, M.~E. and {Tjin A Djie}, H.~R.~E.},
        title = "{The stellar composition of the star formation region CMa R1 - I. Results from new photometric and spectroscopic classifications}",
      journal = {MNRAS},
         year = 1999,
        month = nov,
       volume = {310},
       number = {1},
        pages = {210-222},
          doi = {10.1046/j.1365-8711.1999.02937.x},
       adsurl = {https://ui.adsabs.harvard.edu/abs/1999MNRAS.310..210S}
}

@ARTICLE{Fernandes2019,
       author = {{Fernandes}, B. and {Montmerle}, T. and {Santos-Silva}, T. and {Gregorio-Hetem}, J.},
        title = "{Runaways and shells around the CMa OB1 association}",
      journal = {A\&A},
         year = 2019,
        month = aug,
       volume = {628},
          eid = {A44},
        pages = {A44},
          doi = {10.1051/0004-6361/201935484},
archivePrefix = {arXiv},
       eprint = {1906.00113},
 primaryClass = {astro-ph.SR},
       adsurl = {https://ui.adsabs.harvard.edu/abs/2019A&A...628A..44F}
}

@ARTICLE{Fischer2016,
       author = {{Fischer}, William J. and {Padgett}, Deborah L. and {Stapelfeldt}, Karl L. and {Sewi{\l}o}, Marta},
        title = "{A WISE Census of Young Stellar Objects in Canis Major}",
      journal = {ApJ},
         year = 2016,
        month = aug,
       volume = {827},
       number = {2},
          eid = {96},
        pages = {96},
          doi = {10.3847/0004-637X/827/2/96},
archivePrefix = {arXiv},
       eprint = {1606.01896},
 primaryClass = {astro-ph.SR},
       adsurl = {https://ui.adsabs.harvard.edu/abs/2016ApJ...827...96F}
}

@ARTICLE{Alcala2002,
       author = {{Alcal{\'a}}, J.~M. and {Covino}, E. and {Melo}, C. and {Sterzik}, M.~F.},
        title = "{Characterization of low-mass pre-main sequence stars in the Southern Cross}",
      journal = {A\&A},
         year = 2002,
        month = mar,
       volume = {384},
        pages = {521-531},
          doi = {10.1051/0004-6361:20020070},
       adsurl = {https://ui.adsabs.harvard.edu/abs/2002A&A...384..521A}
}

@ARTICLE{Marigo2017,
       author = {{Marigo}, Paola and {Girardi}, L{\'e}o and {Bressan}, Alessandro and {Rosenfield}, Philip and {Aringer}, Bernhard and {Chen}, Yang and {Dussin}, Marco and {Nanni}, Ambra and {Pastorelli}, Giada and {Rodrigues}, Tha{\'\i}se S. and {Trabucchi}, Michele and {Bladh}, Sara and {Dalcanton}, Julianne and {Groenewegen}, Martin A.~T. and {Montalb{\'a}n}, Josefina and {Wood}, Peter R.},
        title = "{A New Generation of PARSEC-COLIBRI Stellar Isochrones Including the TP-AGB Phase}",
      journal = {ApJ},
         year = 2017,
        month = jan,
       volume = {835},
       number = {1},
          eid = {77},
        pages = {77},
          doi = {10.3847/1538-4357/835/1/77},
archivePrefix = {arXiv},
       eprint = {1701.08510},
 primaryClass = {astro-ph.SR},
       adsurl = {https://ui.adsabs.harvard.edu/abs/2017ApJ...835...77M}
}

@ARTICLE{Claes2024,
       author = {{Claes}, R.~A.~B. and {Campbell-White}, J. and {Manara}, C.~F. and {Frasca}, A. and {Natta}, A. and {Alcal{\'a}}, J.~M. and {Armeni}, A. and {Fang}, M. and {Lovell}, J.~B. and {Stelzer}, B. and {Venuti}, L. and {Wyatt}, M. and {Queitsch}, A.},
        title = "{FitteR for Accretion ProPErties of T Tauri stars (FRAPPE): A new approach to use class III spectra to derive stellar and accretion properties}",
      journal = {\aap},
         year = 2024,
        month = oct,
       volume = {690},
          eid = {A122},
        pages = {A122},
          doi = {10.1051/0004-6361/202450885},
archivePrefix = {arXiv},
       eprint = {2407.11866},
 primaryClass = {astro-ph.SR},
       adsurl = {https://ui.adsabs.harvard.edu/abs/2024A&A...690A.122C},
}

@ARTICLE{Gordon2023,
       author = {{Gordon}, Karl D. and {Clayton}, Geoffrey C. and {Decleir}, Marjorie and {Fitzpatrick}, E.~L. and {Massa}, Derck and {Misselt}, Karl A. and {Tollerud}, Erik J.},
        title = "{One Relation for All Wavelengths: The Far-ultraviolet to Mid-infrared Milky Way Spectroscopic R(V)-dependent Dust Extinction Relationship}",
      journal = {\apj},
         year = 2023,
        month = jun,
       volume = {950},
       number = {2},
          eid = {86},
        pages = {86},
          doi = {10.3847/1538-4357/accb59},
archivePrefix = {arXiv},
       eprint = {2304.01991},
 primaryClass = {astro-ph.GA},
       adsurl = {https://ui.adsabs.harvard.edu/abs/2023ApJ...950...86G}
}

@ARTICLE{Fitzpatrick1999,
       author = {{Fitzpatrick}, Edward L.},
        title = "{Correcting for the Effects of Interstellar Extinction}",
      journal = {\pasp},
         year = 1999,
        month = jan,
       volume = {111},
       number = {755},
        pages = {63-75},
          doi = {10.1086/316293},
archivePrefix = {arXiv},
       eprint = {astro-ph/9809387},
 primaryClass = {astro-ph},
       adsurl = {https://ui.adsabs.harvard.edu/abs/1999PASP..111...63F}
}

@ARTICLE{Herczeg2014,
       author = {{Herczeg}, Gregory J. and {Hillenbrand}, Lynne A.},
        title = "{An Optical Spectroscopic Study of T Tauri Stars. I. Photospheric Properties}",
      journal = {\apj},
         year = 2014,
        month = may,
       volume = {786},
       number = {2},
          eid = {97},
        pages = {97},
          doi = {10.1088/0004-637X/786/2/97},
archivePrefix = {arXiv},
       eprint = {1403.1675},
 primaryClass = {astro-ph.SR},
       adsurl = {https://ui.adsabs.harvard.edu/abs/2014ApJ...786...97H}
}

@ARTICLE{Jeffries2007,
       author = {{Jeffries}, R.~D. and {Oliveira}, J.~M. and {Naylor}, Tim and {Mayne}, N.~J. and {Littlefair}, S.~P.},
        title = "{The Keele-Exeter young cluster survey - I. Low-mass pre-main-sequence stars in NGC 2169}",
      journal = {\mnras},
         year = 2007,
        month = apr,
       volume = {376},
       number = {2},
        pages = {580-598},
          doi = {10.1111/j.1365-2966.2007.11327.x},
archivePrefix = {arXiv},
       eprint = {astro-ph/0611630},
 primaryClass = {astro-ph},
       adsurl = {https://ui.adsabs.harvard.edu/abs/2007MNRAS.376..580J}
}

@ARTICLE{Andrae2023,
       author = {{Andrae}, R. and {Fouesneau}, M. and {Sordo}, R. and {Bailer-Jones}, C.~A.~L. and {Dharmawardena}, T.~E. and {Rybizki}, J. and {De Angeli}, F. and {Lindstr{\o}m}, H.~E.~P. and {Marshall}, D.~J. and {Drimmel}, R. and {Korn}, A.~J. and {Soubiran}, C. and {Brouillet}, N. and {Casamiquela}, L. and {Rix}, H. -W. and {Abreu Aramburu}, A. and {{\'A}lvarez}, M.~A. and {Bakker}, J. and {Bellas-Velidis}, I. and {Bijaoui}, A. and {Brugaletta}, E. and {Burlacu}, A. and {Carballo}, R. and {Chaoul}, L. and {Chiavassa}, A. and {Contursi}, G. and {Cooper}, W.~J. and {Creevey}, O.~L. and {Dafonte}, C. and {Dapergolas}, A. and {de Laverny}, P. and {Delchambre}, L. and {Demouchy}, C. and {Edvardsson}, B. and {Fr{\'e}mat}, Y. and {Garabato}, D. and {Garc{\'\i}a-Lario}, P. and {Garc{\'\i}a-Torres}, M. and {Gavel}, A. and {Gomez}, A. and {Gonz{\'a}lez-Santamar{\'\i}a}, I. and {Hatzidimitriou}, D. and {Heiter}, U. and {Jean-Antoine Piccolo}, A. and {Kontizas}, M. and {Kordopatis}, G. and {Lanzafame}, A.~C. and {Lebreton}, Y. and {Licata}, E.~L. and {Livanou}, E. and {Lobel}, A. and {Lorca}, A. and {Magdaleno Romeo}, A. and {Manteiga}, M. and {Marocco}, F. and {Mary}, N. and {Nicolas}, C. and {Ordenovic}, C. and {Pailler}, F. and {Palicio}, P.~A. and {Pallas-Quintela}, L. and {Panem}, C. and {Pichon}, B. and {Poggio}, E. and {Recio-Blanco}, A. and {Riclet}, F. and {Robin}, C. and {Santove{\~n}a}, R. and {Sarro}, L.~M. and {Schultheis}, M.~S. and {Segol}, M. and {Silvelo}, A. and {Slezak}, I. and {Smart}, R.~L. and {S{\"u}veges}, M. and {Th{\'e}venin}, F. and {Torralba Elipe}, G. and {Ulla}, A. and {Utrilla}, E. and {Vallenari}, A. and {van Dillen}, E. and {Zhao}, H. and {Zorec}, J.},
        title = "{Gaia Data Release 3. Analysis of the Gaia BP/RP spectra using the General Stellar Parameterizer from Photometry}",
      journal = {\aap},
         year = 2023,
        month = jun,
       volume = {674},
          eid = {A27},
        pages = {A27},
          doi = {10.1051/0004-6361/202243462},
archivePrefix = {arXiv},
       eprint = {2206.06138},
 primaryClass = {astro-ph.SR},
       adsurl = {https://ui.adsabs.harvard.edu/abs/2023A&A...674A..27A}
}

@ARTICLE{Green2019,
       author = {{Green}, Gregory M. and {Schlafly}, Edward and {Zucker}, Catherine and {Speagle}, Joshua S. and {Finkbeiner}, Douglas},
        title = "{A 3D Dust Map Based on Gaia, Pan-STARRS 1, and 2MASS}",
      journal = {\apj},
         year = 2019,
        month = dec,
       volume = {887},
       number = {1},
          eid = {93},
        pages = {93},
          doi = {10.3847/1538-4357/ab5362},
archivePrefix = {arXiv},
       eprint = {1905.02734},
 primaryClass = {astro-ph.GA},
       adsurl = {https://ui.adsabs.harvard.edu/abs/2019ApJ...887...93G}
}

@ARTICLE{Pecaut2013,
       author = {{Pecaut}, Mark J. and {Mamajek}, Eric E.},
        title = "{Intrinsic Colors, Temperatures, and Bolometric Corrections of Pre-main-sequence Stars}",
      journal = {\apjs},
         year = 2013,
        month = sep,
       volume = {208},
       number = {1},
          eid = {9},
        pages = {9},
          doi = {10.1088/0067-0049/208/1/9},
archivePrefix = {arXiv},
       eprint = {1307.2657},
 primaryClass = {astro-ph.SR},
       adsurl = {https://ui.adsabs.harvard.edu/abs/2013ApJS..208....9P}
}

@ARTICLE{Pecaut2016,
       author = {{Pecaut}, Mark J. and {Mamajek}, Eric E.},
        title = "{The star formation history and accretion-disc fraction among the K-type members of the Scorpius-Centaurus OB association}",
      journal = {\mnras},
         year = 2016,
        month = sep,
       volume = {461},
       number = {1},
        pages = {794-815},
          doi = {10.1093/mnras/stw1300},
archivePrefix = {arXiv},
       eprint = {1605.08789},
 primaryClass = {astro-ph.SR},
       adsurl = {https://ui.adsabs.harvard.edu/abs/2016MNRAS.461..794P}
}

@ARTICLE{Ushomirsky1998,
       author = {{Ushomirsky}, Greg and {Matzner}, Christopher D. and {Brown}, Edward F. and {Bildsten}, Lars and {Hilliard}, Vadim G. and {Schroeder}, Peter C.},
        title = "{Light-Element Depletion in Contracting Brown Dwarfs and Pre-Main-Sequence Stars}",
      journal = {\apj},
         year = 1998,
        month = apr,
       volume = {497},
       number = {1},
        pages = {253-266},
          doi = {10.1086/305457},
archivePrefix = {arXiv},
       eprint = {astro-ph/9711099},
 primaryClass = {astro-ph},
       adsurl = {https://ui.adsabs.harvard.edu/abs/1998ApJ...497..253U}
}

@ARTICLE{Jeffries2023,
       author = {{Jeffries}, R.~D. and {Jackson}, R.~J. and {Wright}, Nicholas J. and {Weaver}, G. and {Gilmore}, G. and {Randich}, S. and {Bragaglia}, A. and {Korn}, A.~J. and {Smiljanic}, R. and {Biazzo}, K. and {Casey}, A.~R. and {Frasca}, A. and {Gonneau}, A. and {Guiglion}, G. and {Morbidelli}, L. and {Prisinzano}, L. and {Sacco}, G.~G. and {Tautvai{\v{s}}ien{\.{e}}}, G. and {Worley}, C.~C. and {Zaggia}, S.},
        title = "{The Gaia-ESO Survey: empirical estimates of stellar ages from lithium equivalent widths (EAGLES)}",
      journal = {\mnras},
         year = 2023,
        month = jul,
       volume = {523},
       number = {1},
        pages = {802-824},
          doi = {10.1093/mnras/stad1293},
archivePrefix = {arXiv},
       eprint = {2304.12197},
 primaryClass = {astro-ph.SR},
       adsurl = {https://ui.adsabs.harvard.edu/abs/2023MNRAS.523..802J}
}

@ARTICLE{Chen2020,
       author = {{Chen}, Xiaodian and {Wang}, Shu and {Deng}, Licai and {de Grijs}, Richard and {Yang}, Ming and {Tian}, Hao},
        title = "{The Zwicky Transient Facility Catalog of Periodic Variable Stars}",
      journal = {\apjs},
         year = 2020,
        month = jul,
       volume = {249},
       number = {1},
          eid = {18},
        pages = {18},
          doi = {10.3847/1538-4365/ab9cae},
archivePrefix = {arXiv},
       eprint = {2005.08662},
 primaryClass = {astro-ph.SR},
       adsurl = {https://ui.adsabs.harvard.edu/abs/2020ApJS..249...18C}
}

@ARTICLE{Bressan2012,
       author = {{Bressan}, Alessandro and {Marigo}, Paola and {Girardi}, L{\'e}o. and {Salasnich}, Bernardo and {Dal Cero}, Claudia and {Rubele}, Stefano and {Nanni}, Ambra},
        title = "{PARSEC: stellar tracks and isochrones with the PAdova and TRieste Stellar Evolution Code}",
      journal = {\mnras},
         year = 2012,
        month = nov,
       volume = {427},
       number = {1},
        pages = {127-145},
          doi = {10.1111/j.1365-2966.2012.21948.x},
archivePrefix = {arXiv},
       eprint = {1208.4498},
 primaryClass = {astro-ph.SR},
       adsurl = {https://ui.adsabs.harvard.edu/abs/2012MNRAS.427..127B}
}

@ARTICLE{Chen2015,
       author = {{Chen}, Yang and {Bressan}, Alessandro and {Girardi}, L{\'e}o and {Marigo}, Paola and {Kong}, Xu and {Lanza}, Antonio},
        title = "{PARSEC evolutionary tracks of massive stars up to 350 M$_{{\ensuremath{\odot}}}$ at metallicities 0.0001 {\ensuremath{\leq}} Z {\ensuremath{\leq}} 0.04}",
      journal = {\mnras},
         year = 2015,
        month = sep,
       volume = {452},
       number = {1},
        pages = {1068-1080},
          doi = {10.1093/mnras/stv1281},
archivePrefix = {arXiv},
       eprint = {1506.01681},
 primaryClass = {astro-ph.SR},
       adsurl = {https://ui.adsabs.harvard.edu/abs/2015MNRAS.452.1068C}
}

@ARTICLE{Chen2014,
       author = {{Chen}, Yang and {Girardi}, L{\'e}o and {Bressan}, Alessandro and {Marigo}, Paola and {Barbieri}, Mauro and {Kong}, Xu},
        title = "{Improving PARSEC models for very low mass stars}",
      journal = {\mnras},
         year = 2014,
        month = nov,
       volume = {444},
       number = {3},
        pages = {2525-2543},
          doi = {10.1093/mnras/stu1605},
archivePrefix = {arXiv},
       eprint = {1409.0322},
 primaryClass = {astro-ph.SR},
       adsurl = {https://ui.adsabs.harvard.edu/abs/2014MNRAS.444.2525C}
}

@ARTICLE{Tang2014,
       author = {{Tang}, Jing and {Bressan}, Alessandro and {Rosenfield}, Philip and {Slemer}, Alessandra and {Marigo}, Paola and {Girardi}, L{\'e}o and {Bianchi}, Luciana},
        title = "{New PARSEC evolutionary tracks of massive stars at low metallicity: testing canonical stellar evolution in nearby star-forming dwarf galaxies}",
      journal = {\mnras},
         year = 2014,
        month = dec,
       volume = {445},
       number = {4},
        pages = {4287-4305},
          doi = {10.1093/mnras/stu2029},
archivePrefix = {arXiv},
       eprint = {1410.1745},
 primaryClass = {astro-ph.SR},
       adsurl = {https://ui.adsabs.harvard.edu/abs/2014MNRAS.445.4287T}
}

@ARTICLE{Pastorelli2019,
       author = {{Pastorelli}, Giada and {Marigo}, Paola and {Girardi}, L{\'e}o and {Chen}, Yang and {Rubele}, Stefano and {Trabucchi}, Michele and {Aringer}, Bernhard and {Bladh}, Sara and {Bressan}, Alessandro and {Montalb{\'a}n}, Josefina and {Boyer}, Martha L. and {Dalcanton}, Julianne J. and {Eriksson}, Kjell and {Groenewegen}, Martin A.~T. and {H{\"o}fner}, Susanne and {Lebzelter}, Thomas and {Nanni}, Ambra and {Rosenfield}, Philip and {Wood}, Peter R. and {Cioni}, Maria-Rosa L.},
        title = "{Constraining the thermally pulsing asymptotic giant branch phase with resolved stellar populations in the Small Magellanic Cloud}",
      journal = {\mnras},
         year = 2019,
        month = jun,
       volume = {485},
       number = {4},
        pages = {5666-5692},
          doi = {10.1093/mnras/stz725},
archivePrefix = {arXiv},
       eprint = {1903.04499},
 primaryClass = {astro-ph.SR},
       adsurl = {https://ui.adsabs.harvard.edu/abs/2019MNRAS.485.5666P}
}

@ARTICLE{Pastorelli2020,
       author = {{Pastorelli}, Giada and {Marigo}, Paola and {Girardi}, L{\'e}o and {Aringer}, Bernhard and {Chen}, Yang and {Rubele}, Stefano and {Trabucchi}, Michele and {Bladh}, Sara and {Boyer}, Martha L. and {Bressan}, Alessandro and {Dalcanton}, Julianne J. and {Groenewegen}, Martin A.~T. and {Lebzelter}, Thomas and {Mowlavi}, Nami and {Chubb}, Katy L. and {Cioni}, Maria-Rosa L. and {de Grijs}, Richard and {Ivanov}, Valentin D. and {Nanni}, Ambra and {van Loon}, Jacco Th and {Zaggia}, Simone},
        title = "{Constraining the thermally pulsing asymptotic giant branch phase with resolved stellar populations in the Large Magellanic Cloud}",
      journal = {\mnras},
         year = 2020,
        month = nov,
       volume = {498},
       number = {3},
        pages = {3283-3301},
          doi = {10.1093/mnras/staa2565},
archivePrefix = {arXiv},
       eprint = {2008.08595},
 primaryClass = {astro-ph.SR},
       adsurl = {https://ui.adsabs.harvard.edu/abs/2020MNRAS.498.3283P}
}

@ARTICLE{Dong2024,
       author = {{Dong}, Yiwei and {Xu}, Ye and {Hao}, Chaojie and {Li}, Yingjie and {Liu}, Dejian and {Sun}, Yan and {Lin}, Zehao},
        title = "{3D Morphology and Motions of the Canis Major Region from Gaia DR3}",
      journal = {\aj},
         year = 2024,
        month = nov,
       volume = {168},
       number = {5},
          eid = {225},
        pages = {225},
          doi = {10.3847/1538-3881/ad77a8},
archivePrefix = {arXiv},
       eprint = {2409.01670},
 primaryClass = {astro-ph.GA},
       adsurl = {https://ui.adsabs.harvard.edu/abs/2024AJ....168..225D}
}

@ARTICLE{Anderson2014,
       author = {{Anderson}, L.~D. and {Bania}, T.~M. and {Balser}, Dana S. and {Cunningham}, V. and {Wenger}, T.~V. and {Johnstone}, B.~M. and {Armentrout}, W.~P.},
        title = "{The WISE Catalog of Galactic H II Regions}",
      journal = {\apjs},
         year = 2014,
        month = may,
       volume = {212},
       number = {1},
          eid = {1},
        pages = {1},
          doi = {10.1088/0067-0049/212/1/1},
archivePrefix = {arXiv},
       eprint = {1312.6202},
 primaryClass = {astro-ph.GA},
       adsurl = {https://ui.adsabs.harvard.edu/abs/2014ApJS..212....1A}
}

@ARTICLE{Sharpless1959,
       author = {{Sharpless}, Stewart},
        title = "{A Catalogue of H II Regions.}",
      journal = {\apjs},
         year = 1959,
        month = dec,
       volume = {4},
        pages = {257},
          doi = {10.1086/190049},
       adsurl = {https://ui.adsabs.harvard.edu/abs/1959ApJS....4..257S}
}

@ARTICLE{LeBorgne2003,
       author = {{Le Borgne}, J. -F. and {Bruzual}, G. and {Pell{\'o}}, R. and {Lan{\c{c}}on}, A. and {Rocca-Volmerange}, B. and {Sanahuja}, B. and {Schaerer}, D. and {Soubiran}, C. and {V{\'\i}lchez-G{\'o}mez}, R.},
        title = "{STELIB: A library of stellar spectra at R \raisebox{-0.5ex}\textasciitilde 2000}",
      journal = {\aap},
         year = 2003,
        month = may,
       volume = {402},
        pages = {433-442},
          doi = {10.1051/0004-6361:20030243},
archivePrefix = {arXiv},
       eprint = {astro-ph/0302334},
 primaryClass = {astro-ph},
       adsurl = {https://ui.adsabs.harvard.edu/abs/2003A&A...402..433L}
}

@ARTICLE{Paolino2025,
       author = {{P{\'e}rez Paolino}, Facundo and {Bary}, Jeffrey S. and {Hillenbrand}, Lynne A. and {Horner}, Benjamin and {Carvalho}, Adolfo},
        title = "{Spectral Biases, Starspot Morphology, and Dynamo Transitions on the Pre-main Sequence: Insights from the X-Shooter WTTS Library}",
      journal = {\apj},
         year = 2025,
        month = sep,
       volume = {990},
       number = {2},
          eid = {205},
        pages = {205},
          doi = {10.3847/1538-4357/adf6ad},
archivePrefix = {arXiv},
       eprint = {2505.10837},
 primaryClass = {astro-ph.SR},
       adsurl = {https://ui.adsabs.harvard.edu/abs/2025ApJ...990..205P}
}

@ARTICLE{Weaver2024,
       author = {{Weaver}, G. and {Jeffries}, R.~D. and {Jackson}, R.~J.},
        title = "{Using neural network models to estimate stellar ages from lithium equivalent widths: an EAGLES expansion}",
      journal = {\mnras},
         year = 2024,
        month = nov,
       volume = {534},
       number = {3},
        pages = {2014-2029},
          doi = {10.1093/mnras/stae2133},
archivePrefix = {arXiv},
       eprint = {2409.07523},
 primaryClass = {astro-ph.SR},
       adsurl = {https://ui.adsabs.harvard.edu/abs/2024MNRAS.534.2014W}
}

@ARTICLE{GaiaDR32023,
       author = {{Gaia Collaboration} and {Vallenari}, A. and {Brown}, A.~G.~A. and {Prusti}, T. and {de Bruijne}, J.~H.~J. and {Arenou}, F. and {Babusiaux}, C. and {Biermann}, M. and {Creevey}, O.~L. and {Ducourant}, C. and {Evans}, D.~W. and {Eyer}, L. and {Guerra}, R. and {Hutton}, A. and {Jordi}, C. and {Klioner}, S.~A. and {Lammers}, U.~L. and {Lindegren}, L. and {Luri}, X. and {Mignard}, F. and {Panem}, C. and {Pourbaix}, D. and {Randich}, S. and {Sartoretti}, P. and {Soubiran}, C. and {Tanga}, P. and {Walton}, N.~A. and {Bailer-Jones}, C.~A.~L. and {Bastian}, U. and {Drimmel}, R. and {Jansen}, F. and {Katz}, D. and {Lattanzi}, M.~G. and {van Leeuwen}, F. and {Bakker}, J. and {Cacciari}, C. and {Casta{\~n}eda}, J. and {De Angeli}, F. and {Fabricius}, C. and {Fouesneau}, M. and {Fr{\'e}mat}, Y. and {Galluccio}, L. and {Guerrier}, A. and {Heiter}, U. and {Masana}, E. and {Messineo}, R. and {Mowlavi}, N. and {Nicolas}, C. and {Nienartowicz}, K. and {Pailler}, F. and {Panuzzo}, P. and {Riclet}, F. and {Roux}, W. and {Seabroke}, G.~M. and {Sordo}, R. and {Th{\'e}venin}, F. and {Gracia-Abril}, G. and {Portell}, J. and {Teyssier}, D. and {Altmann}, M. and {Andrae}, R. and {Audard}, M. and {Bellas-Velidis}, I. and {Benson}, K. and {Berthier}, J. and {Blomme}, R. and {Burgess}, P.~W. and {Busonero}, D. and {Busso}, G. and {C{\'a}novas}, H. and {Carry}, B. and {Cellino}, A. and {Cheek}, N. and {Clementini}, G. and {Damerdji}, Y. and {Davidson}, M. and {de Teodoro}, P. and {Nu{\~n}ez Campos}, M. and {Delchambre}, L. and {Dell'Oro}, A. and {Esquej}, P. and {Fern{\'a}ndez-Hern{\'a}ndez}, J. and {Fraile}, E. and {Garabato}, D. and {Garc{\'\i}a-Lario}, P. and {Gosset}, E. and {Haigron}, R. and {Halbwachs}, J. -L. and {Hambly}, N.~C. and {Harrison}, D.~L. and {Hern{\'a}ndez}, J. and {Hestroffer}, D. and {Hodgkin}, S.~T. and {Holl}, B. and {Jan{\ss}en}, K. and {Jevardat de Fombelle}, G. and {Jordan}, S. and {Krone-Martins}, A. and {Lanzafame}, A.~C. and {L{\"o}ffler}, W. and {Marchal}, O. and {Marrese}, P.~M. and {Moitinho}, A. and {Muinonen}, K. and {Osborne}, P. and {Pancino}, E. and {Pauwels}, T. and {Recio-Blanco}, A. and {Reyl{\'e}}, C. and {Riello}, M. and {Rimoldini}, L. and {Roegiers}, T. and {Rybizki}, J. and {Sarro}, L.~M. and {Siopis}, C. and {Smith}, M. and {Sozzetti}, A. and {Utrilla}, E. and {van Leeuwen}, M. and {Abbas}, U. and {{\'A}brah{\'a}m}, P. and {Abreu Aramburu}, A. and {Aerts}, C. and {Aguado}, J.~J. and {Ajaj}, M. and {Aldea-Montero}, F. and {Altavilla}, G. and {{\'A}lvarez}, M.~A. and {Alves}, J. and {Anders}, F. and {Anderson}, R.~I. and {Anglada Varela}, E. and {Antoja}, T. and {Baines}, D. and {Baker}, S.~G. and {Balaguer-N{\'u}{\~n}ez}, L. and {Balbinot}, E. and {Balog}, Z. and {Barache}, C. and {Barbato}, D. and {Barros}, M. and {Barstow}, M.~A. and {Bartolom{\'e}}, S. and {Bassilana}, J. -L. and {Bauchet}, N. and {Becciani}, U. and {Bellazzini}, M. and {Berihuete}, A. and {Bernet}, M. and {Bertone}, S. and {Bianchi}, L. and {Binnenfeld}, A. and {Blanco-Cuaresma}, S. and {Blazere}, A. and {Boch}, T. and {Bombrun}, A. and {Bossini}, D. and {Bouquillon}, S. and {Bragaglia}, A. and {Bramante}, L. and {Breedt}, E. and {Bressan}, A. and {Brouillet}, N. and {Brugaletta}, E. and {Bucciarelli}, B. and {Burlacu}, A. and {Butkevich}, A.~G. and {Buzzi}, R. and {Caffau}, E. and {Cancelliere}, R. and {Cantat-Gaudin}, T. and {Carballo}, R. and {Carlucci}, T. and {Carnerero}, M.~I. and {Carrasco}, J.~M. and {Casamiquela}, L. and {Castellani}, M. and {Castro-Ginard}, A. and {Chaoul}, L. and {Charlot}, P. and {Chemin}, L. and {Chiaramida}, V. and {Chiavassa}, A. and {Chornay}, N. and {Comoretto}, G. and {Contursi}, G. and {Cooper}, W.~J. and {Cornez}, T. and {Cowell}, S. and {Crifo}, F. and {Cropper}, M. and {Crosta}, M. and {Crowley}, C. and {Dafonte}, C. and {Dapergolas}, A. and {David}, M. and {David}, P. and {de Laverny}, P. and {De Luise}, F. and {De March}, R.},
        title = "{Gaia Data Release 3. Summary of the content and survey properties}",
      journal = {\aap},
         year = 2023,
        month = jun,
       volume = {674},
          eid = {A1},
        pages = {A1},
          doi = {10.1051/0004-6361/202243940},
archivePrefix = {arXiv},
       eprint = {2208.00211},
 primaryClass = {astro-ph.GA},
       adsurl = {https://ui.adsabs.harvard.edu/abs/2023A&A...674A...1G}
}

@INPROCEEDINGS{Soderblom2014,
       author = {{Soderblom}, D.~R. and {Hillenbrand}, L.~A. and {Jeffries}, R.~D. and {Mamajek}, E.~E. and {Naylor}, T.},
        title = "{Ages of Young Stars}",
    booktitle = {Protostars and Planets VI},
         year = 2014,
       editor = {{Beuther}, Henrik and {Klessen}, Ralf S. and {Dullemond}, Cornelis P. and {Henning}, Thomas},
        month = jan,
        pages = {219-241},
          doi = {10.2458/azu_uapress_9780816531240-ch010},
archivePrefix = {arXiv},
       eprint = {1311.7024},
 primaryClass = {astro-ph.SR},
       adsurl = {https://ui.adsabs.harvard.edu/abs/2014prpl.conf..219S}
}

@ARTICLE{Odonnell1994,
       author = {{O'Donnell}, James E.},
        title = "{R v-dependent Optical and Near-Ultraviolet Extinction}",
      journal = {\apj},
         year = 1994,
        month = feb,
       volume = {422},
        pages = {158},
          doi = {10.1086/173713},
       adsurl = {https://ui.adsabs.harvard.edu/abs/1994ApJ...422..158O}
}

@ARTICLE{Cardelli1989,
       author = {{Cardelli}, Jason A. and {Clayton}, Geoffrey C. and {Mathis}, John S.},
        title = "{The Relationship between Infrared, Optical, and Ultraviolet Extinction}",
      journal = {\apj},
         year = 1989,
        month = oct,
       volume = {345},
        pages = {245},
          doi = {10.1086/167900},
       adsurl = {https://ui.adsabs.harvard.edu/abs/1989ApJ...345..245C}
}

@ARTICLE{Comeron1998,
       author = {{Comeron}, F. and {Torra}, J. and {Gomez}, A.~E.},
        title = "{Kinematic signatures of violent formation of galactic OB associations from HIPPARCOS measurements}",
      journal = {\aap},
         year = 1998,
        month = feb,
       volume = {330},
        pages = {975-989},
       adsurl = {https://ui.adsabs.harvard.edu/abs/1998A&A...330..975C}
}

@ARTICLE{He2022,
       author = {{He}, Zhihong and {Wang}, Kun and {Luo}, Yangping and {Li}, Jing and {Liu}, Xiaochen and {Jiang}, Qingquan},
        title = "{A Blind All-sky Search for Star Clusters in Gaia EDR3: 886 Clusters within 1.2 kpc of the Sun}",
      journal = {\apjs},
         year = 2022,
        month = sep,
       volume = {262},
       number = {1},
          eid = {7},
        pages = {7},
          doi = {10.3847/1538-4365/ac7c17},
archivePrefix = {arXiv},
       eprint = {2206.12170},
 primaryClass = {astro-ph.GA},
       adsurl = {https://ui.adsabs.harvard.edu/abs/2022ApJS..262....7H}
}

@ARTICLE{Delfini2025,
       author = {{Delfini}, L. and {Vioque}, M. and {Ribas}, {\'A}. and {Hodgkin}, S.},
        title = "{Star formation and accretion rates within 500 pc as traced by Gaia DR3 XP spectra}",
      journal = {\aap},
         year = 2025,
        month = jul,
       volume = {699},
          eid = {A145},
        pages = {A145},
          doi = {10.1051/0004-6361/202453539},
archivePrefix = {arXiv},
       eprint = {2505.04699},
 primaryClass = {astro-ph.SR},
       adsurl = {https://ui.adsabs.harvard.edu/abs/2025A&A...699A.145D}
}

@ARTICLE{Green2018_dustmaps,
       author = {{Green}, Gregory M.},
        title = "{dustmaps: A Python interface for maps of interstellar dust}",
      journal = {The Journal of Open Source Software},
         year = 2018,
        month = jun,
       volume = {3},
       number = {26},
        pages = {695},
          doi = {10.21105/joss.00695},
       adsurl = {https://ui.adsabs.harvard.edu/abs/2018JOSS....3..695G}
}

@INPROCEEDINGS{Tody1986_IRAF,
       author = {{Tody}, Doug},
        title = "{The IRAF Data Reduction and Analysis System}",
    booktitle = {Instrumentation in astronomy VI},
         year = 1986,
       editor = {{Crawford}, David L.},
       series = {Society of Photo-Optical Instrumentation Engineers (SPIE) Conference Series},
       volume = {627},
        month = jan,
        pages = {733},
          doi = {10.1117/12.968154},
       adsurl = {https://ui.adsabs.harvard.edu/abs/1986SPIE..627..733T}
}

@INPROCEEDINGS{Taylor2005_TOPCAT,
       author = {{Taylor}, M.~B.},
        title = "{TOPCAT \& STIL: Starlink Table/VOTable Processing Software}",
    booktitle = {Astronomical Data Analysis Software and Systems XIV},
         year = 2005,
       editor = {{Shopbell}, P. and {Britton}, M. and {Ebert}, R.},
       series = {Astronomical Society of the Pacific Conference Series},
       volume = {347},
        month = dec,
        pages = {29},
       adsurl = {https://ui.adsabs.harvard.edu/abs/2005ASPC..347...29T}
}

@INPROCEEDINGS{Taylor2006_STILTS,
       author = {{Taylor}, M.~B.},
        title = "{STILTS - A Package for Command-Line Processing of Tabular Data}",
    booktitle = {Astronomical Data Analysis Software and Systems XV},
         year = 2006,
       editor = {{Gabriel}, C. and {Arviset}, C. and {Ponz}, D. and {Enrique}, S.},
       series = {Astronomical Society of the Pacific Conference Series},
       volume = {351},
        month = jul,
        pages = {666},
       adsurl = {https://ui.adsabs.harvard.edu/abs/2006ASPC..351..666T}
}

@Article{harris2020_numpy,
 title         = {Array programming with {NumPy}},
 author        = {Charles R. Harris and K. Jarrod Millman and St{\'{e}}fan J.
                 van der Walt and Ralf Gommers and Pauli Virtanen and David
                 Cournapeau and Eric Wieser and Julian Taylor and Sebastian
                 Berg and Nathaniel J. Smith and Robert Kern and Matti Picus
                 and Stephan Hoyer and Marten H. van Kerkwijk and Matthew
                 Brett and Allan Haldane and Jaime Fern{\'{a}}ndez del
                 R{\'{i}}o and Mark Wiebe and Pearu Peterson and Pierre
                 G{\'{e}}rard-Marchant and Kevin Sheppard and Tyler Reddy and
                 Warren Weckesser and Hameer Abbasi and Christoph Gohlke and
                 Travis E. Oliphant},
 year          = {2020},
 month         = sep,
 journal       = {Nature},
 volume        = {585},
 number        = {7825},
 pages         = {357--362},
 doi           = {10.1038/s41586-020-2649-2},
 publisher     = {Springer Science and Business Media {LLC}},
 url           = {https://doi.org/10.1038/s41586-020-2649-2}
}

@InProceedings{ mckinney-proc-scipy-2010_PANDAS,
  author    = { {W}es {M}c{K}inney },
  title     = { {D}ata {S}tructures for {S}tatistical {C}omputing in {P}ython },
  booktitle = { {P}roceedings of the 9th {P}ython in {S}cience {C}onference },
  pages     = { 56 - 61 },
  year      = { 2010 },
  editor    = { {S}t\'efan van der {W}alt and {J}arrod {M}illman },
  doi       = { 10.25080/Majora-92bf1922-00a }
}

@Article{Hunter2007_Matplotlib,
  Author    = {Hunter, J. D.},
  Title     = {Matplotlib: A 2D graphics environment},
  Journal   = {Computing in Science \& Engineering},
  Volume    = {9},
  Number    = {3},
  Pages     = {90--95},
  publisher = {IEEE COMPUTER SOC},
  doi       = {10.1109/MCSE.2007.55},
  year      = 2007
}

@ARTICLE{2020SciPy-NMeth,
  author  = {Virtanen, Pauli and Gommers, Ralf and Oliphant, Travis E. and
            Haberland, Matt and Reddy, Tyler and Cournapeau, David and
            Burovski, Evgeni and Peterson, Pearu and Weckesser, Warren and
            Bright, Jonathan and {van der Walt}, St{\'e}fan J. and
            Brett, Matthew and Wilson, Joshua and Millman, K. Jarrod and
            Mayorov, Nikolay and Nelson, Andrew R. J. and Jones, Eric and
            Kern, Robert and Larson, Eric and Carey, C J and
            Polat, {\.I}lhan and Feng, Yu and Moore, Eric W. and
            {VanderPlas}, Jake and Laxalde, Denis and Perktold, Josef and
            Cimrman, Robert and Henriksen, Ian and Quintero, E. A. and
            Harris, Charles R. and Archibald, Anne M. and
            Ribeiro, Ant{\^o}nio H. and Pedregosa, Fabian and
            {van Mulbregt}, Paul and {SciPy 1.0 Contributors}},
  title   = {{{SciPy} 1.0: Fundamental Algorithms for Scientific
            Computing in Python}},
  journal = {Nature Methods},
  year    = {2020},
  volume  = {17},
  pages   = {261--272},
  adsurl  = {https://rdcu.be/b08Wh},
  doi     = {10.1038/s41592-019-0686-2},
}

@ARTICLE{2022Astropy1,
       author = {{Astropy Collaboration} and {Price-Whelan}, Adrian M. and {Lim}, Pey Lian and {Earl}, Nicholas and {Starkman}, Nathaniel and {Bradley}, Larry and {Shupe}, David L. and {Patil}, Aarya A. and {Corrales}, Lia and {Brasseur}, C.~E. and {N{\"o}the}, Maximilian and {Donath}, Axel and {Tollerud}, Erik and {Morris}, Brett M. and {Ginsburg}, Adam and {Vaher}, Eero and {Weaver}, Benjamin A. and {Tocknell}, James and {Jamieson}, William and {van Kerkwijk}, Marten H. and {Robitaille}, Thomas P. and {Merry}, Bruce and {Bachetti}, Matteo and {G{\"u}nther}, H. Moritz and {Aldcroft}, Thomas L. and {Alvarado-Montes}, Jaime A. and {Archibald}, Anne M. and {B{\'o}di}, Attila and {Bapat}, Shreyas and {Barentsen}, Geert and {Baz{\'a}n}, Juanjo and {Biswas}, Manish and {Boquien}, M{\'e}d{\'e}ric and {Burke}, D.~J. and {Cara}, Daria and {Cara}, Mihai and {Conroy}, Kyle E. and {Conseil}, Simon and {Craig}, Matthew W. and {Cross}, Robert M. and {Cruz}, Kelle L. and {D'Eugenio}, Francesco and {Dencheva}, Nadia and {Devillepoix}, Hadrien A.~R. and {Dietrich}, J{\"o}rg P. and {Eigenbrot}, Arthur Davis and {Erben}, Thomas and {Ferreira}, Leonardo and {Foreman-Mackey}, Daniel and {Fox}, Ryan and {Freij}, Nabil and {Garg}, Suyog and {Geda}, Robel and {Glattly}, Lauren and {Gondhalekar}, Yash and {Gordon}, Karl D. and {Grant}, David and {Greenfield}, Perry and {Groener}, Austen M. and {Guest}, Steve and {Gurovich}, Sebastian and {Handberg}, Rasmus and {Hart}, Akeem and {Hatfield-Dodds}, Zac and {Homeier}, Derek and {Hosseinzadeh}, Griffin and {Jenness}, Tim and {Jones}, Craig K. and {Joseph}, Prajwel and {Kalmbach}, J. Bryce and {Karamehmetoglu}, Emir and {Ka{\l}uszy{\'n}ski}, Miko{\l}aj and {Kelley}, Michael S.~P. and {Kern}, Nicholas and {Kerzendorf}, Wolfgang E. and {Koch}, Eric W. and {Kulumani}, Shankar and {Lee}, Antony and {Ly}, Chun and {Ma}, Zhiyuan and {MacBride}, Conor and {Maljaars}, Jakob M. and {Muna}, Demitri and {Murphy}, N.~A. and {Norman}, Henrik and {O'Steen}, Richard and {Oman}, Kyle A. and {Pacifici}, Camilla and {Pascual}, Sergio and {Pascual-Granado}, J. and {Patil}, Rohit R. and {Perren}, Gabriel I. and {Pickering}, Timothy E. and {Rastogi}, Tanuj and {Roulston}, Benjamin R. and {Ryan}, Daniel F. and {Rykoff}, Eli S. and {Sabater}, Jose and {Sakurikar}, Parikshit and {Salgado}, Jes{\'u}s and {Sanghi}, Aniket and {Saunders}, Nicholas and {Savchenko}, Volodymyr and {Schwardt}, Ludwig and {Seifert-Eckert}, Michael and {Shih}, Albert Y. and {Jain}, Anany Shrey and {Shukla}, Gyanendra and {Sick}, Jonathan and {Simpson}, Chris and {Singanamalla}, Sudheesh and {Singer}, Leo P. and {Singhal}, Jaladh and {Sinha}, Manodeep and {Sip{\H{o}}cz}, Brigitta M. and {Spitler}, Lee R. and {Stansby}, David and {Streicher}, Ole and {{\v{S}}umak}, Jani and {Swinbank}, John D. and {Taranu}, Dan S. and {Tewary}, Nikita and {Tremblay}, Grant R. and {de Val-Borro}, Miguel and {Van Kooten}, Samuel J. and {Vasovi{\'c}}, Zlatan and {Verma}, Shresth and {de Miranda Cardoso}, Jos{\'e} Vin{\'\i}cius and {Williams}, Peter K.~G. and {Wilson}, Tom J. and {Winkel}, Benjamin and {Wood-Vasey}, W.~M. and {Xue}, Rui and {Yoachim}, Peter and {Zhang}, Chen and {Zonca}, Andrea and {Astropy Project Contributors}},
        title = "{The Astropy Project: Sustaining and Growing a Community-oriented Open-source Project and the Latest Major Release (v5.0) of the Core Package}",
      journal = {\apj},
         year = 2022,
        month = aug,
       volume = {935},
       number = {2},
          eid = {167},
        pages = {167},
          doi = {10.3847/1538-4357/ac7c74},
archivePrefix = {arXiv},
       eprint = {2206.14220},
 primaryClass = {astro-ph.IM},
       adsurl = {https://ui.adsabs.harvard.edu/abs/2022ApJ...935..167A}
}

@ARTICLE{2018Astropy2,
       author = {{Astropy Collaboration} and {Price-Whelan}, A.~M. and {Sip{\H{o}}cz}, B.~M. and {G{\"u}nther}, H.~M. and {Lim}, P.~L. and {Crawford}, S.~M. and {Conseil}, S. and {Shupe}, D.~L. and {Craig}, M.~W. and {Dencheva}, N. and {Ginsburg}, A. and {VanderPlas}, J.~T. and {Bradley}, L.~D. and {P{\'e}rez-Su{\'a}rez}, D. and {de Val-Borro}, M. and {Aldcroft}, T.~L. and {Cruz}, K.~L. and {Robitaille}, T.~P. and {Tollerud}, E.~J. and {Ardelean}, C. and {Babej}, T. and {Bach}, Y.~P. and {Bachetti}, M. and {Bakanov}, A.~V. and {Bamford}, S.~P. and {Barentsen}, G. and {Barmby}, P. and {Baumbach}, A. and {Berry}, K.~L. and {Biscani}, F. and {Boquien}, M. and {Bostroem}, K.~A. and {Bouma}, L.~G. and {Brammer}, G.~B. and {Bray}, E.~M. and {Breytenbach}, H. and {Buddelmeijer}, H. and {Burke}, D.~J. and {Calderone}, G. and {Cano Rodr{\'\i}guez}, J.~L. and {Cara}, M. and {Cardoso}, J.~V.~M. and {Cheedella}, S. and {Copin}, Y. and {Corrales}, L. and {Crichton}, D. and {D'Avella}, D. and {Deil}, C. and {Depagne}, {\'E}. and {Dietrich}, J.~P. and {Donath}, A. and {Droettboom}, M. and {Earl}, N. and {Erben}, T. and {Fabbro}, S. and {Ferreira}, L.~A. and {Finethy}, T. and {Fox}, R.~T. and {Garrison}, L.~H. and {Gibbons}, S.~L.~J. and {Goldstein}, D.~A. and {Gommers}, R. and {Greco}, J.~P. and {Greenfield}, P. and {Groener}, A.~M. and {Grollier}, F. and {Hagen}, A. and {Hirst}, P. and {Homeier}, D. and {Horton}, A.~J. and {Hosseinzadeh}, G. and {Hu}, L. and {Hunkeler}, J.~S. and {Ivezi{\'c}}, {\v{Z}}. and {Jain}, A. and {Jenness}, T. and {Kanarek}, G. and {Kendrew}, S. and {Kern}, N.~S. and {Kerzendorf}, W.~E. and {Khvalko}, A. and {King}, J. and {Kirkby}, D. and {Kulkarni}, A.~M. and {Kumar}, A. and {Lee}, A. and {Lenz}, D. and {Littlefair}, S.~P. and {Ma}, Z. and {Macleod}, D.~M. and {Mastropietro}, M. and {McCully}, C. and {Montagnac}, S. and {Morris}, B.~M. and {Mueller}, M. and {Mumford}, S.~J. and {Muna}, D. and {Murphy}, N.~A. and {Nelson}, S. and {Nguyen}, G.~H. and {Ninan}, J.~P. and {N{\"o}the}, M. and {Ogaz}, S. and {Oh}, S. and {Parejko}, J.~K. and {Parley}, N. and {Pascual}, S. and {Patil}, R. and {Patil}, A.~A. and {Plunkett}, A.~L. and {Prochaska}, J.~X. and {Rastogi}, T. and {Reddy Janga}, V. and {Sabater}, J. and {Sakurikar}, P. and {Seifert}, M. and {Sherbert}, L.~E. and {Sherwood-Taylor}, H. and {Shih}, A.~Y. and {Sick}, J. and {Silbiger}, M.~T. and {Singanamalla}, S. and {Singer}, L.~P. and {Sladen}, P.~H. and {Sooley}, K.~A. and {Sornarajah}, S. and {Streicher}, O. and {Teuben}, P. and {Thomas}, S.~W. and {Tremblay}, G.~R. and {Turner}, J.~E.~H. and {Terr{\'o}n}, V. and {van Kerkwijk}, M.~H. and {de la Vega}, A. and {Watkins}, L.~L. and {Weaver}, B.~A. and {Whitmore}, J.~B. and {Woillez}, J. and {Zabalza}, V. and {Astropy Contributors}},
        title = "{The Astropy Project: Building an Open-science Project and Status of the v2.0 Core Package}",
      journal = {\aj},
         year = 2018,
        month = sep,
       volume = {156},
       number = {3},
          eid = {123},
        pages = {123},
          doi = {10.3847/1538-3881/aabc4f},
archivePrefix = {arXiv},
       eprint = {1801.02634},
 primaryClass = {astro-ph.IM},
       adsurl = {https://ui.adsabs.harvard.edu/abs/2018AJ....156..123A}
}

@ARTICLE{2013Astropy3,
       author = {{Astropy Collaboration} and {Robitaille}, Thomas P. and {Tollerud}, Erik J. and {Greenfield}, Perry and {Droettboom}, Michael and {Bray}, Erik and {Aldcroft}, Tom and {Davis}, Matt and {Ginsburg}, Adam and {Price-Whelan}, Adrian M. and {Kerzendorf}, Wolfgang E. and {Conley}, Alexander and {Crighton}, Neil and {Barbary}, Kyle and {Muna}, Demitri and {Ferguson}, Henry and {Grollier}, Fr{\'e}d{\'e}ric and {Parikh}, Madhura M. and {Nair}, Prasanth H. and {Unther}, Hans M. and {Deil}, Christoph and {Woillez}, Julien and {Conseil}, Simon and {Kramer}, Roban and {Turner}, James E.~H. and {Singer}, Leo and {Fox}, Ryan and {Weaver}, Benjamin A. and {Zabalza}, Victor and {Edwards}, Zachary I. and {Azalee Bostroem}, K. and {Burke}, D.~J. and {Casey}, Andrew R. and {Crawford}, Steven M. and {Dencheva}, Nadia and {Ely}, Justin and {Jenness}, Tim and {Labrie}, Kathleen and {Lim}, Pey Lian and {Pierfederici}, Francesco and {Pontzen}, Andrew and {Ptak}, Andy and {Refsdal}, Brian and {Servillat}, Mathieu and {Streicher}, Ole},
        title = "{Astropy: A community Python package for astronomy}",
      journal = {\aap},
         year = 2013,
        month = oct,
       volume = {558},
          eid = {A33},
        pages = {A33},
          doi = {10.1051/0004-6361/201322068},
archivePrefix = {arXiv},
       eprint = {1307.6212},
 primaryClass = {astro-ph.IM},
       adsurl = {https://ui.adsabs.harvard.edu/abs/2013A&A...558A..33A}
}

@ARTICLE{Merino2022,
       author = {{Merino}, B. and {Placco}, V. and {Stanghellini}, L.},
        title = "{The US NGO GMOS Data Reduction Cookbook: Version 2.0}",
      journal = {The NOIRLab Mirror},
         year = 2022,
        month = jun,
       volume = {3},
        pages = {13},
       adsurl = {https://ui.adsabs.harvard.edu/abs/2022Mirro...3...13M}
}

@software{nicholas_earl_2025_SPECUTILS,
  author       = {Nicholas Earl and
                  Erik Tollerud and
                  Ricky O'Steen and
                  brechmos and
                  Wolfgang Kerzendorf and
                  Ivo Busko and
                  shaileshahuja and
                  P. L. Lim and
                  Dan D'Avella and
                  Thomas Robitaille and
                  Adam Ginsburg and
                  Derek Homeier and
                  Brigitta Sipőcz and
                  Jesse Averbukh and
                  Brian Cherinka and
                  James Tocknell and
                  Sara Ogaz and
                  Robel Geda and
                  James Davies and
                  Kyle Conroy and
                  Hans Moritz Günther and
                  Kyle Barbary and
                  Kelle Cruz and
                  Jonathan Foster and
                  Michael Droettboom and
                  Duy Nguyen and
                  E. M. Bray and
                  Andy Casey and
                  Henry Ferguson},
  title        = {astropy/specutils: v2.1.0},
  month        = jul,
  year         = 2025,
  publisher    = {Zenodo},
  version      = {v2.1.0},
  doi          = {10.5281/zenodo.16615456},
  url          = {https://doi.org/10.5281/zenodo.16615456},
}

@ARTICLE{2024Gordon_dust_extinction,
       author = {{Gordon}, Karl},
        title = "{dust\_extinction: Interstellar Dust Extinction Models}",
      journal = {The Journal of Open Source Software},
         year = 2024,
        month = aug,
       volume = {9},
       number = {100},
          eid = {7023},
        pages = {7023},
          doi = {10.21105/joss.07023},
       adsurl = {https://ui.adsabs.harvard.edu/abs/2024JOSS....9.7023G}
}

@ARTICLE{Dobashi2013,
       author = {{Dobashi}, Kazuhito and {Marshall}, Douglas J. and {Shimoikura}, Tomomi and {Bernard}, Jean-Philippe},
        title = "{Atlas and Catalog of Dark Clouds Based on the 2 Micron All Sky Survey. II. Correction of the Background Using the Besan{\c{c}}on Galaxy Model}",
      journal = {\pasj},
         year = 2013,
        month = apr,
       volume = {65},
          eid = {31},
        pages = {31},
          doi = {10.1093/pasj/65.2.31},
       adsurl = {https://ui.adsabs.harvard.edu/abs/2013PASJ...65...31D}
}

\begin{appendix}

\onecolumn

\FloatBarrier

\section{Spectra of the TTs in the GMOS sample} \label{ap: spectra}
\FloatBarrier

Here we present portions of the spectra of the remaining 28 TTs not shown previously in the main text, focussing on the spectral regions around H$\alpha$, Li~I ($\lambda~6708$~\AA) and Ca~I ($\lambda$ 6718~\AA) features.

   \begin{figure}[htb!]
   \centering
   \includegraphics[width=0.45\hsize]{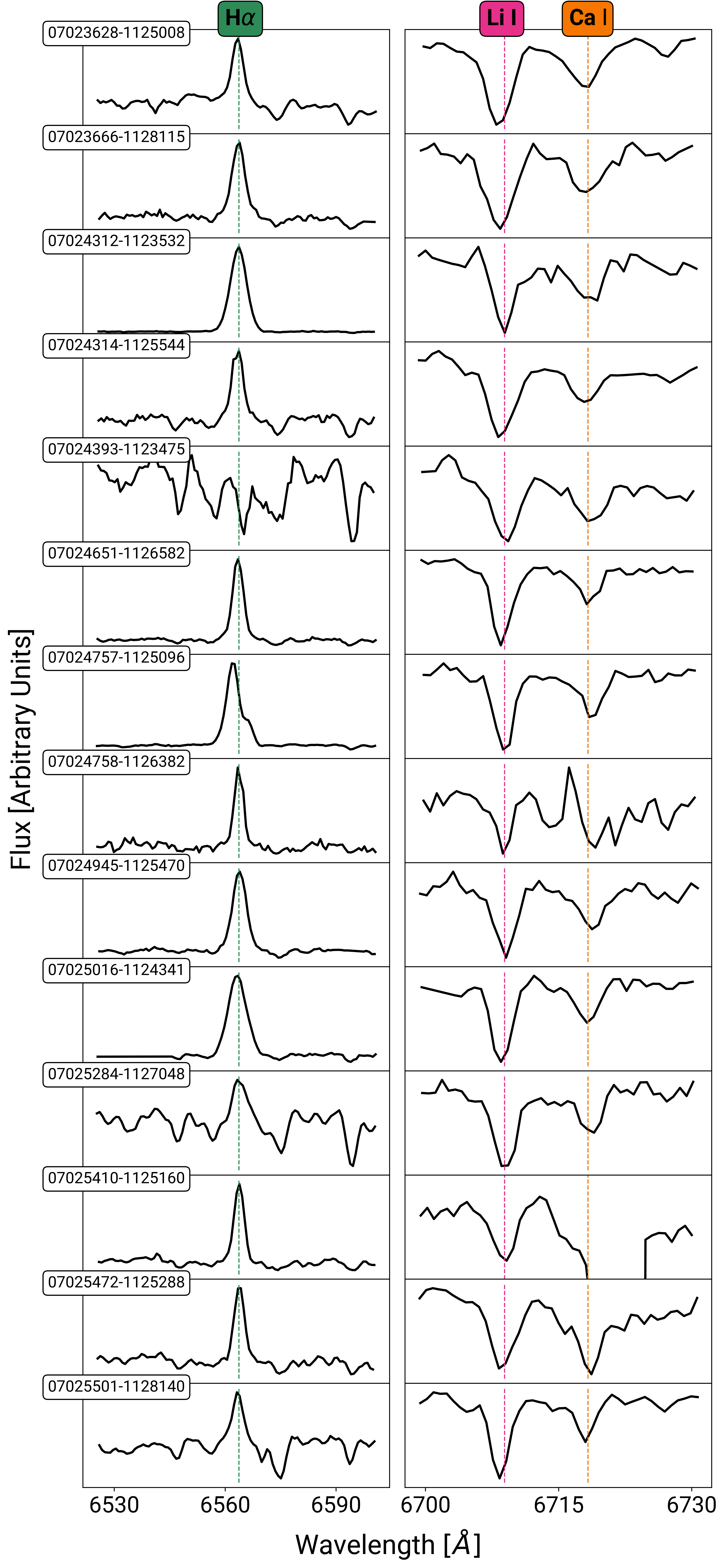}
   \includegraphics[width=0.45\hsize]{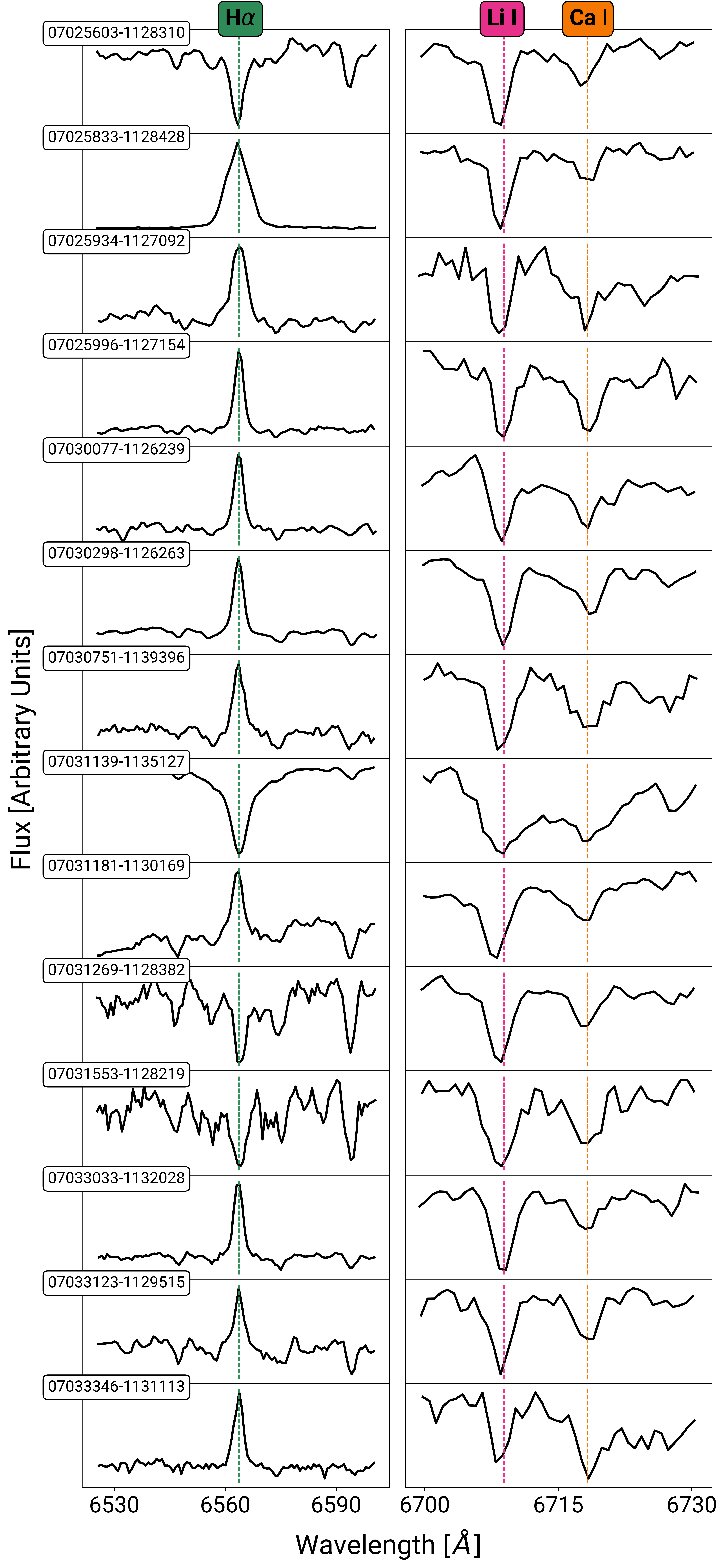}
      \caption{GMOS spectra of 28 TTs identified in this work. The H$\alpha$, Li~I ($\lambda~6708$~\AA) and Ca~I ($\lambda$~6718~\AA) features are highlighted. Each object is labelled with its 2MASS identifier. An artefact is present near the Ca~I line in the spectrum of 2MASS J07025410-1125160.}
     \label{fig: specs}
   \end{figure}

\FloatBarrier

\section{List of additional H$\alpha$ emitters} \label{ap: HalphaEmm}
\FloatBarrier

In this appendix, we present a list of objects with H$\alpha$ emission, but without detectable Li~I absorption. The six sources found are shown in Table~\ref{tab: Ha_em}. One of them does not have an infrared counterpart in the 2MASS catalogue, considering 1\arcsec~maximum separation.

None of them are in the lists of members previously determined by our team \citep{Santos-Silva2021,gregorio-hetem2021b}, and only one was previously found in the survey for H$\alpha$ emission conducted by \citet{Pettersson2019} --- object 07031171-1135369, which also shows the highest H$\alpha$ emission among these six objects. These sources may  be  dMe stars; however, further investigation is required.

\begin{table*}[htb!]
\caption{H$\alpha$ emitters without Li absorption line. }
\label{tab: Ha_em}
\centering 
\begin{tabular}{l c c c c}
\hline\hline 
2MASS & H$\alpha$ & $G_{Gaia}$ & $G_{BP}$ & $G_{RP}$\\
 & (\AA) & (mag) & (mag) & (mag) \\
\hline

07024699-1127121 & -2.137 & 18.83 & 20.23 & 17.58 \\
07025060-1125390 & -4.818 & 19.33 & 21.03 & 18.06 \\
07031171-1135369 & -10.129 & 16.20 & 17.24 & 15.17 \\
07031992-1131462 & -4.606 & 19.79 & 20.96 & 18.39 \\
07032216-1133071 & -3.398 & 19.93 & 21.53 & 18.39 \\
(07033274-1131121)\tablefootmark{a} & -2.135 & 20.23 & 21.26 & 19.05 \\
07033410-1132587 & -5.962 & 19.32 & 20.72 & 18.16 \\

\hline
\end{tabular}
\tablefoot{\tablefoottext{a}{Object without 2MASS counterpart for which we adopted an identification following the same nomenclature.}
}
\end{table*}

\FloatBarrier
\section{Extinction} \label{ap: ext}
\FloatBarrier

\begin{figure}[htb!]
   \centering
   \includegraphics[width=1\hsize]{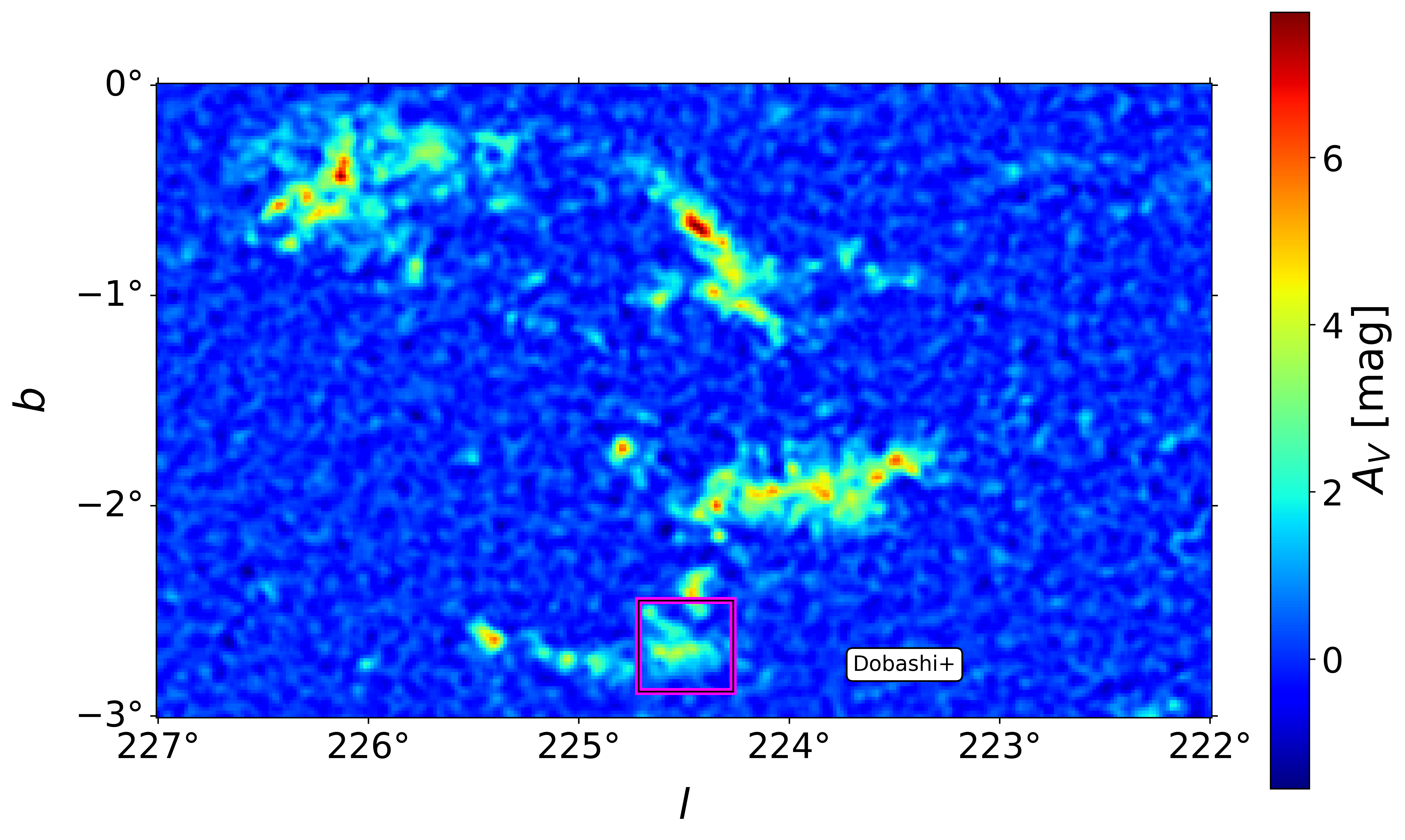}
   \includegraphics[width=1\hsize]{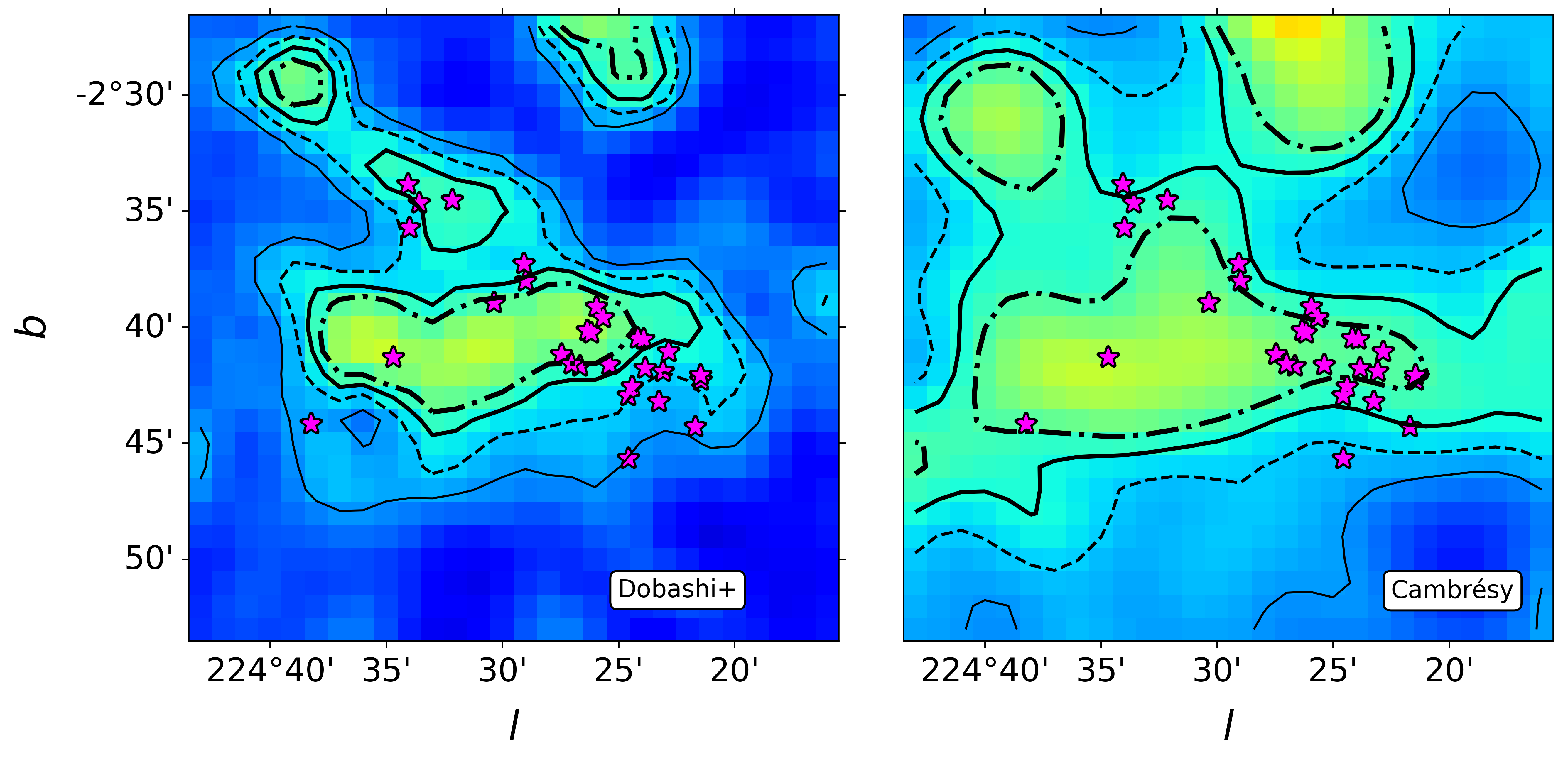}
      \caption{2D extinction maps from Dobashi+ and Cambrésy. The top panel indicates the CMa region while the bottom panel shows a zoom in the region highlighted in magenta in the top panel. Magenta star symbols indicate the position of all 29 TT stars studied in this work.}
         \label{fig: extinction_maps}
   \end{figure}

We compared our extinction estimates, obtained from spectral fitting, with estimates from two-dimensional maps: one from Cambrésy (private communication), and another one from \citet{Dobashi2011} and \citet{Dobashi2013}, hereafter Dobashi+ (Fig.~\ref{fig: extinction_maps}). The resolution of these maps varies between 1~\arcmin and 12~\arcmin. We also used a three-dimensional map \citep[Bayestar19 --- ][]{Green2019} with a typical scale of 3\arcmin4 – 13\arcmin7, as well as  the estimates from {\it Gaia} DR3 --- General Stellar Parametrizer from Photometry \citep[GSP-Phot,][]{Andrae2023}. 

\begin{figure}[htb!]
   \centering
   \includegraphics[width=0.61\hsize]{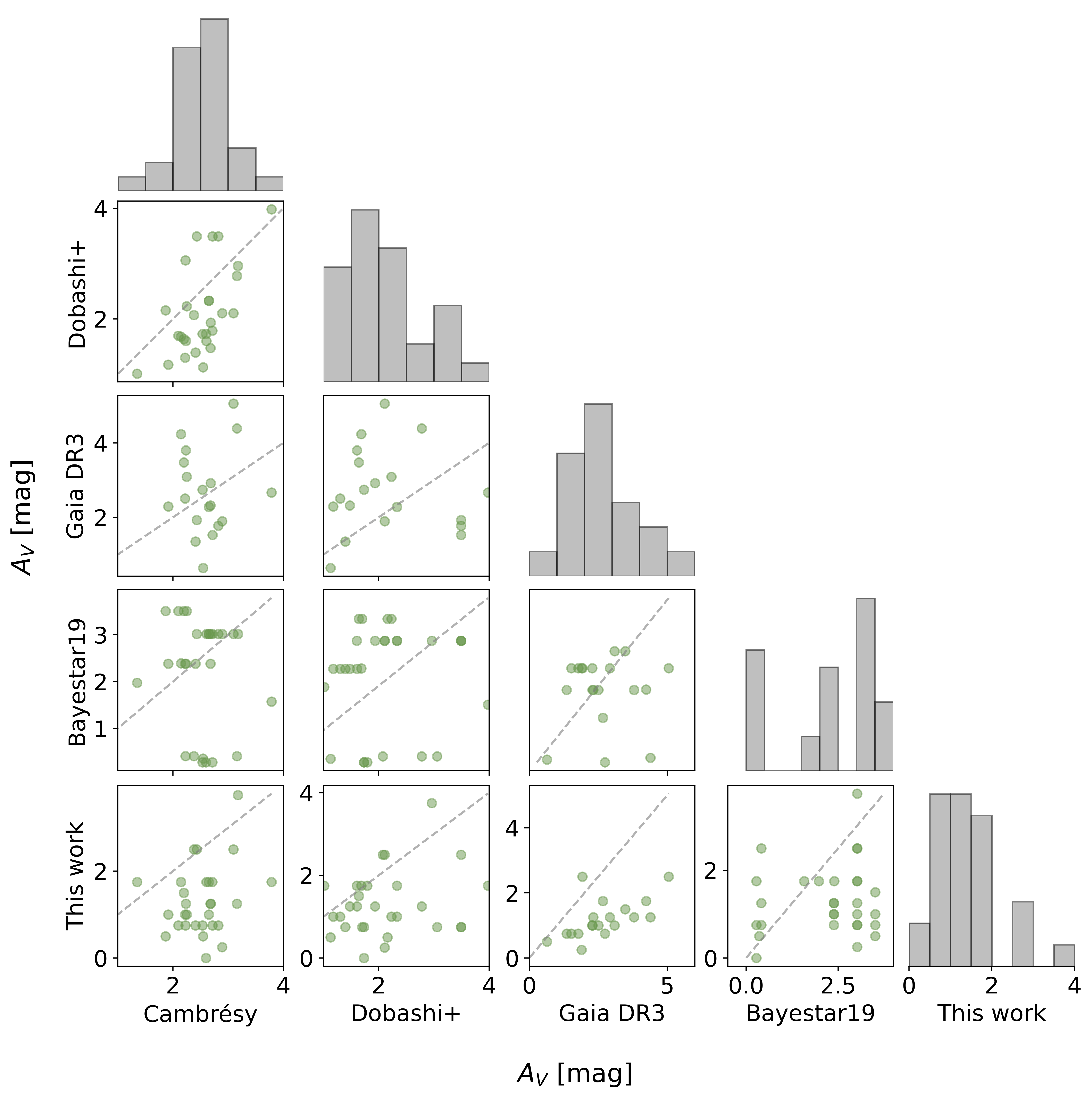}
      \caption{Comparison of extinction estimates for the TTs identified in this work with values from the literature.}
         \label{fig: extinction}
   \end{figure}

None of the literature extinction values showed good agreement, either among themselves or with our values (Fig. \ref{fig: extinction}). The Spearman correlation coefficient between the distributions typically ranges from -0.08 to 0.2, indicating very weak or negligible correlations. The only exceptions are between this work and GSP-Phot, which yields a higher correlation (Spearman $\sim$ 0.7), and between Dobashi+ and Cambrésy (Spearman $\sim$ 0.6). Although our extinction values were rounded to 0.25~mag steps, this does not seem to explain the discrepancies. One possibility is that since the 2D and 3D maps cannot resolve individual stars, they do not represent the exact line-of-sight values for each star. In particular, the 2D maps do not take into account the distance to each object, which is important to objects that are not deeply embedded in the molecular clouds. Considering that the $A_V$ values of this work tend to be lower than the $A_V$ maps, we suggest that our sample is mainly found in the borders of the clouds, as can be noted in Fig~\ref{fig: extinction_maps}. On the other hand, GSP-Phot does not account for different physical process that happens in PMS, such as accretion. There is also a degeneracy between GSP-Phot temperature and extinction estimates \citep{Andrae2023} which is another possible reason for the differences found.

\begin{figure}[htb!]
   \centering
   \includegraphics[width=0.61\hsize]{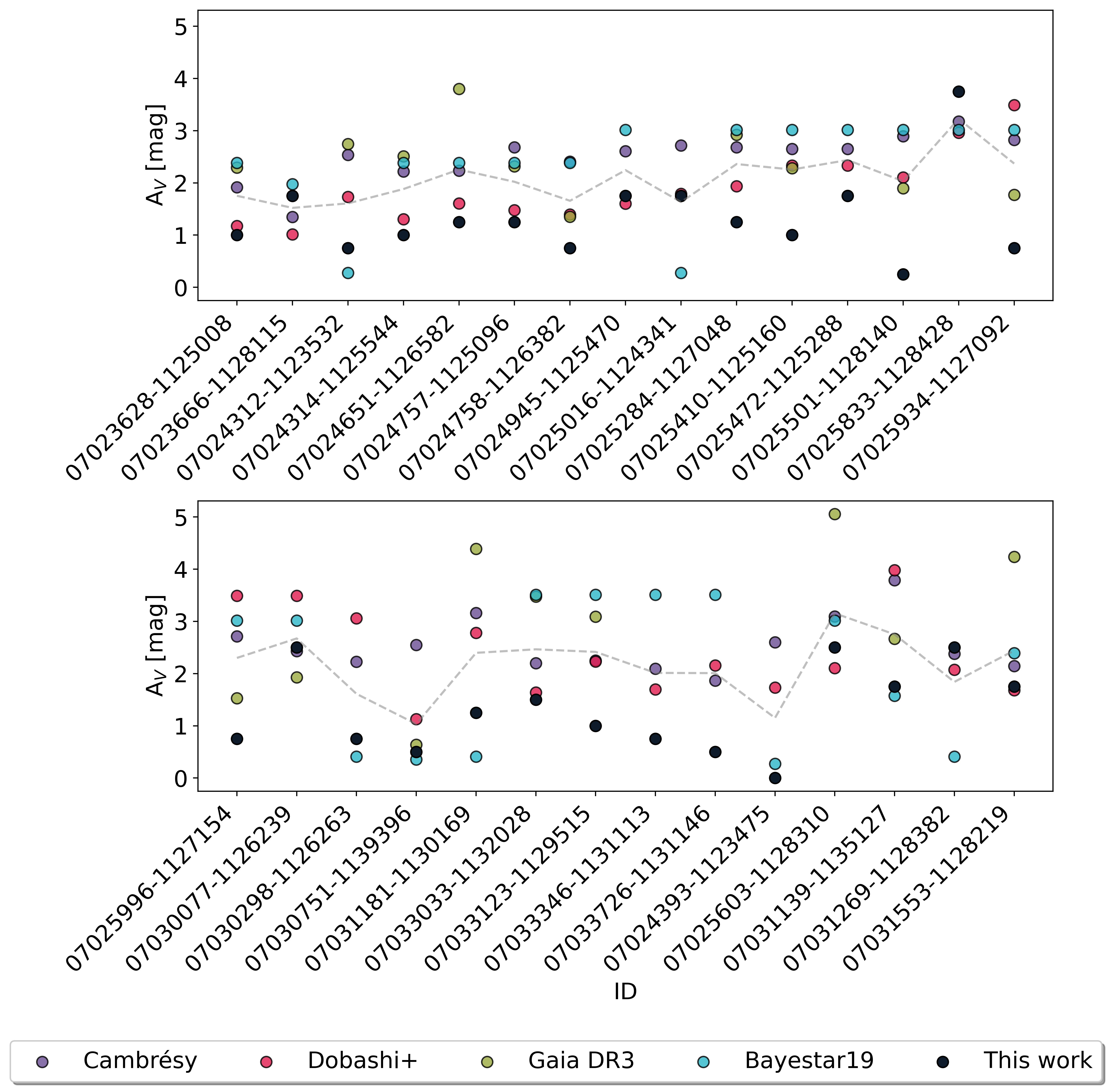}
      \caption{Visual extinction estimates for each TTs (identified by the 2MASS id) from different methods.}
         \label{fig: ext_stars}
   \end{figure}

Figure~\ref{fig: ext_stars} shows the amplitude of the extinction for each TTs in our sample. For approximately  55\% of the sources, our estimates are lower than all the literature values. Conversely, our values are  very conservative in most cases. Notably, the extinction distribution adopted here provides the best agreement with theoretical models in the CMD (Fig.~\ref{fig: CMD}). These results indicate that poor agreement among the extinction estimates is mainly due to differences in the adopted methods rather than wrong spectral type allocation.

\FloatBarrier

\section{Spectral type estimates using different methodologies} \label{ap: teff_estimates}
\FloatBarrier

Table \ref{tab: spec_tp} lists the spectral types derived for each source using different methods (mentioned in Sect. \ref{sec: spt_type}) along with the final adopted values.  

\begin{table*}[htb!]
\caption{Different estimates of spectral type.}
\label{tab: spec_tp}
\centering 
\begin{tabular}{l c c c c c c} 
\hline\hline 
2MASS & Adopted & L18 & C24 & TiO (J07) & TiO 6800 & TiO 7140 \\
\hline

07023628-1125008 & K7 & K7 & G5 & K5.8 & K6/6.5 & $\cdots$ \\
07023666-1128115 & K7.5 & M0 & K7.5 & K8 & K7.5 & $\cdots$ \\
07024312-1123532 & K7 & M0 & K7 & K5.7 & K7.1 & $\cdots$ \\
07024314-1125544 & K7.5 & M0 & K7.5 & K6 & K7 & $\cdots$ \\
07024393-1123475 & K7 & K7 & K2 & $\cdots$ & K6.1 & $\cdots$ \\
07024651-1126582 & K7.5 & M0 & K7.5 & K6.6/K7.2 & K7.5 & $\cdots$ \\
07024757-1125096 & K7 & K7 & G5 & K5.2/5.4 & K6.5 & $\cdots$ \\
07024758-1126382 & M2 & M2.5 & M2 & M0.9/M1.4 & $\cdots$ & M1.2/M1.6 \\
07024945-1125470 & K7.5 & M0.5 & K7.5 & K8 & K7.3 & $\cdots$ \\
07025016-1124341 & K7 & K7 & G5 & K6.3 & K6.3 & $\cdots$ \\
07025284-1127048 & K2 & K2V & K2 & K5 & $\cdots$ & $\cdots$ \\
07025410-1125160 & M1.5 & M1.5 & M1 & K6/M0.6 & M0.2 & M1 \\
07025472-1125288 & K7.5 & M1.5 & K7.5 & M0 & K6.8/K8.3 & M0.5 \\
07025501-1128140 & K6 & K7 & K2/G & K5 & K5.6/6.5 & $\cdots$ \\
07025603-1128310 & K2 & K2V & K2 & $\cdots$ & K5.3 & $\cdots$ \\
07025833-1128428 & K0.5 & K0V & K0.5 & K5.4 & K6.1 & $\cdots$ \\
07025934-1127092 & M2 & M2.5 & M2 & M1.7 & $\cdots$ & M2 \\
07025996-1127154 & M2 & M2.5 & M2 & M1/M2.3 & $\cdots$ & M1.4/M2.7 \\
07030077-1126239 & K7.5 & M1.5 & K7.5 & M0 & K8 & M0.5 \\
07030298-1126263 & K7.5 & M0 & K7.5 & K7.1 & K6 & $\cdots$ \\
07030751-1139396 & K7.5 & M0.5/K8 & K7.5 & K6.5 & K8.8 & $\cdots$ \\
07031139-1135127 & G3 & G3Va & G5 & $\cdots$ & K5.7 & $\cdots$ \\
07031181-1130169 & K7 & K7 & K0.5 & K5.5/K7.3 & K6.1 & $\cdots$ \\
07031269-1128382 & K0.5 & K2V & K0.5 & $\cdots$ & K6 & $\cdots$ \\
07031553-1128219 & K4 & K5 & K4 & K5.2 & K6.3 & $\cdots$ \\
07033033-1132028 & K7.5 & M0 & K7.5 & K5.5 & K6.9 & $\cdots$ \\
07033123-1129515 & K7 & K7 & K7 & K5.5 & K6.5 & $\cdots$ \\
07033346-1131113 & M3.5 & M3.5 & M3.5 & M2.3 & $\cdots$ & M2.3 \\
07033726-1131146 & M0.5 & M0.5-M1.5 & M0.5 & K8.2 & M0 & $\cdots$ \\

\hline

\end{tabular}
\end{table*}

\FloatBarrier 
\twocolumn

\end{appendix}
\end{document}